\documentclass[twocolumn]{aastex631}
\usepackage{natbib}

\newcommand{\kms}{\mathrm{km\,s^{-1}}}
\newcommand{\au}{\mathrm{au}}
\newcommand{\yr}{\mathrm{yr^{-1}}}
\newcommand{\msun}{\mathrm{M_\odot}}
\newcommand{\mstar}{\mathrm{M_*}}
\newcommand{\rsun}{\mathrm{R_\odot}}
\newcommand{\rstar}{\mathrm{R_{*}}}
\newcommand{\lsun}{\mathrm{L_\odot}}

\begin{document}

\title{Multiple shells driven by disk winds\\
ALMA observations in the HH 30 outflow}
\shortauthors{L\'opez-V\'azquez et al.}

\author[0000-0002-5845-8722]{J. A. L\'opez-V\'azquez}
\thanks{Corresponding author: J. A. L\'opez-V\'azquez}
\email{jlopezv@asiaa.sinica.edu.tw}
\affiliation{Academia Sinica Institute of Astronomy and Astrophysics, No. 1, Sec. 4, Roosevelt Road, Taipei 10617, Taiwan}

\author[0000-0002-3024-5864]{Chin-Fei Lee}
\affiliation{Academia Sinica Institute of Astronomy and Astrophysics, No. 1, Sec. 4, Roosevelt Road, Taipei 10617, Taiwan}

\author[0000-0001-5811-0454]{M. Fern\'andez-L\'opez}
\affiliation{Instituto Argentino de Radioastronom\'ia (CCT-La Plata, CONICET; CICPBA), C.C. No. 5, 1894, Villa Elisa, Buenos Aires, Argentina}

\author[0000-0003-3814-4424]{Fabien Louvet}
\affiliation{Univ. Grenoble Alpes, CNRS, IPAG, 38000 Grenoble, France}
\affiliation{Departamento de Astronom\'ia, Universidad de Chile, Casilla 36-D, Santiago ,Chile}

\author[0000-0003-4753-8759]{O. Guerra-Alvarado}
\affiliation{Leiden Observatory, Leiden University, PO Box 9513, 2300 RA Leiden, The Netherlands}

\author[0000-0003-2343-7937]{Luis A. Zapata}
\affiliation{Instituto de Radioastronom\'ia y Astrof\'isica, Universidad Nacional Aut\'onoma de M\'exico, Apartado Postal 3-72, 58089 Morelia, Michoac\'an, M\'exico}



\begin{abstract}
We present archive Atacama Large Millimeter/Submillimeter Array (ALMA) Band 6 observations of the $^{13}$CO (J=2--1) and $^{12}$CO (J=2--1) molecular line emission of the protostellar system associated with HH 30. The $^{13}$CO molecular line shows the accretion disk while the molecular outflow is traced by the emission of the $^{12}$CO molecular line. We estimated a dynamical mass for the central object of $0.45\pm0.14\,\msun$, and a mass for the molecular outflow of $1.83\pm0.19\times10^{-4}\,\msun$. The molecular outflow presents an internal cavity as well as multiple outflowing shell structures. We distinguish three different shells with constant expansion ($\sim4-6\,\kms$) and possible rotation signatures ($\leq0.5\,\kms$). We find that the shells can be explained by magnetocentrifugal disk winds with launching radii $R_\mathrm{launch}\lesssim4\,\au$ and a small magnetic lever arm $\lambda\sim1.6-1.9$. The multiple shell structure may be the result of episodic ejections of the material from the accretion disk associated with three different epochs with dynamical ages of $497\pm15$ yr, $310\pm9$ yr, and $262\pm11$ yr for the first, second, and third shells, respectively. The outermost shell was ejected $187\pm17$ yr before the medium shell, while the medium shell was launched $48\pm14$ yr before the innermost shell. Our estimations of the linear and angular momentum rates of the outflow as well as the accretion luminosity are consistent with the expected values if the outflow of HH 30 is produced by a wide-angle disk wind.
\end{abstract}

\keywords{Accretion (14) -- Herbig-Haro objects (722) -- Star formation (1569) -- Stellar winds (1636) -- Young stellar objects (1834)}


\section{Introduction} \label{sec:introduction}

The molecular outflows and protostellar jets are phenomena present in the star formation process. However, the link between these flows and the connection with the protostar-disk system, are still open questions. These flows play an important role in the evolution of the protostar-disk system because they could be responsible for extracting the excess of angular momentum and limit the mass from the protostar-disk system (\citealt{Blandford1982} and \citealt{Shu1993}).

The protostellar jets are explained as winds ejected directly from inner regions of the accretion disk, very close to the central protostar, by the magnetocentrifugal mechanism where the magnetic field anchored to the accretion disk drives and collimates them (see reviews by \citealt{Konigl2000}; \citealt{Shu2000}). The molecular outflows are interpreted in two different ways: as swept-up material or as material directly ejected from the disk. In the former interpretation, the swept-up gas traces the interaction between the protostellar jet (or a slow disk-wind) and the infalling envelope or parent core (\citealt{Zhang2016}). This interpretation has been used to explain Class 0 and I systems (e.g., \citealt{Lee2000}), as well as very massive molecular outflows (e.g., \citealt{Zapata2015}; \citealt{JALV2019}; \citealt{JALV2020}). 
In the alternative explanation, the molecular outflows comprise material directly from the accretion disk (e.g., \citealt{Pudritz1986}). This interpretation can explain the rotation signatures found in several sources, such as CB 26 (\citealt{Launhardt2009}; \citealt{JALV2023}), Ori-S6 (\citealt{Zapata2010}), HH 797 (\citealt{Pech2012}), TMC1A (\citealt{Bjerkeli2016}), Orion Source I (\citealt{Hirota2017}), HH 212 (\citealt{Lee2018}), HH 30 (\citealt{Louvet2018}), NGC 1333 IRAS 4C (\citealt{Zhang2018}), DG Tau B (\citealt{deValon2020}; \citealt{deValon2022}), and HD 163296 (\citealt{Booth2021}).

For the very massive molecular outflows associated with DG Tau B (\citealt{Zapata2015}) and Orion Source I (\citealt{JALV2020}), the authors presented that the slow disk-winds ejected directly from the accretion disk do not have enough mass, thus these winds cannot account for the observed linear and angular momentum rates of these outflows. Their argument is based on the assumption that the mass-loss rate of the wind is a small fraction of the disk mass accretion rate. Therefore, these large masses of molecular outflows could be explained if the outflow is formed by entrained material from the parent cloud.
 Nevertheless, the estimated rates at millimeter wavelengths are a lower limit. For a more realistic estimate of the mass, linear, and angular momentum rates, it is necessary to consider the emission of the different atomic and molecular lines of the bipolar outflow detected in the optical and near-infrared wavelengths, as shown in several sources such as HH 211 (\citealt{Ray2023}), and B335, HOPS153, HOPS370, IRAS16253, and IRAS20126 (\citealt{Federman2023}). For the case of DG Tau B, the linear momentum rate of the outflow is similar to the rate measured in the high-velocity atomic jet (\citealt{Mitchell1994}; \citealt{Podio2011}), although, if we consider the contribution of the jet, the large discrepancy between the rates are not explained.
Recent non-ideal magnetohydrodynamic simulations of the disk-wind show that this fraction could be large enough to explain large masses (e.g., \citealt{Bai2017}; \citealt{Wang2019}). Also, the abundances of the molecules used for the mass estimation can vary between one or two orders of magnitude (\citealt{Wright2022}), hence the H$_2$ mass in these outflows could be much lower than estimated by the outflows of DG Tau B and Orion Source I. However, both mechanisms may act simultaneously and the jet and wide-angle wind could coexist.

\begin{figure*}[t!]
\centering
\includegraphics[scale=0.475]{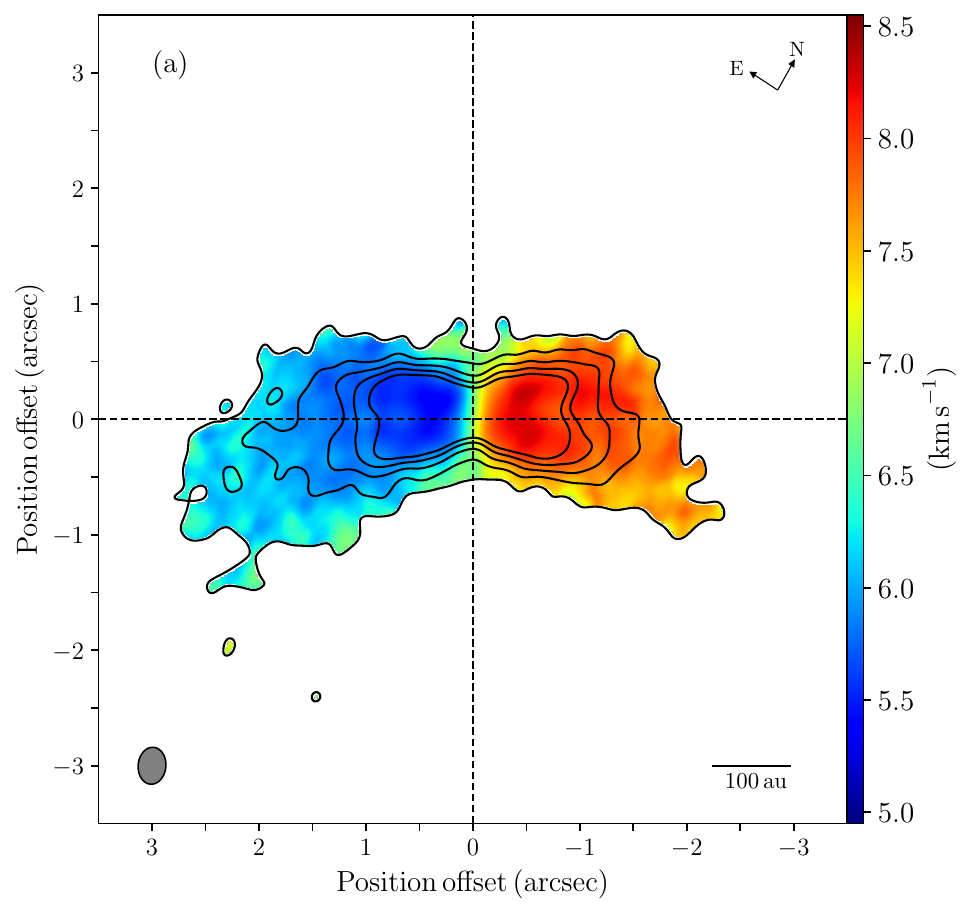}\includegraphics[scale=0.475]{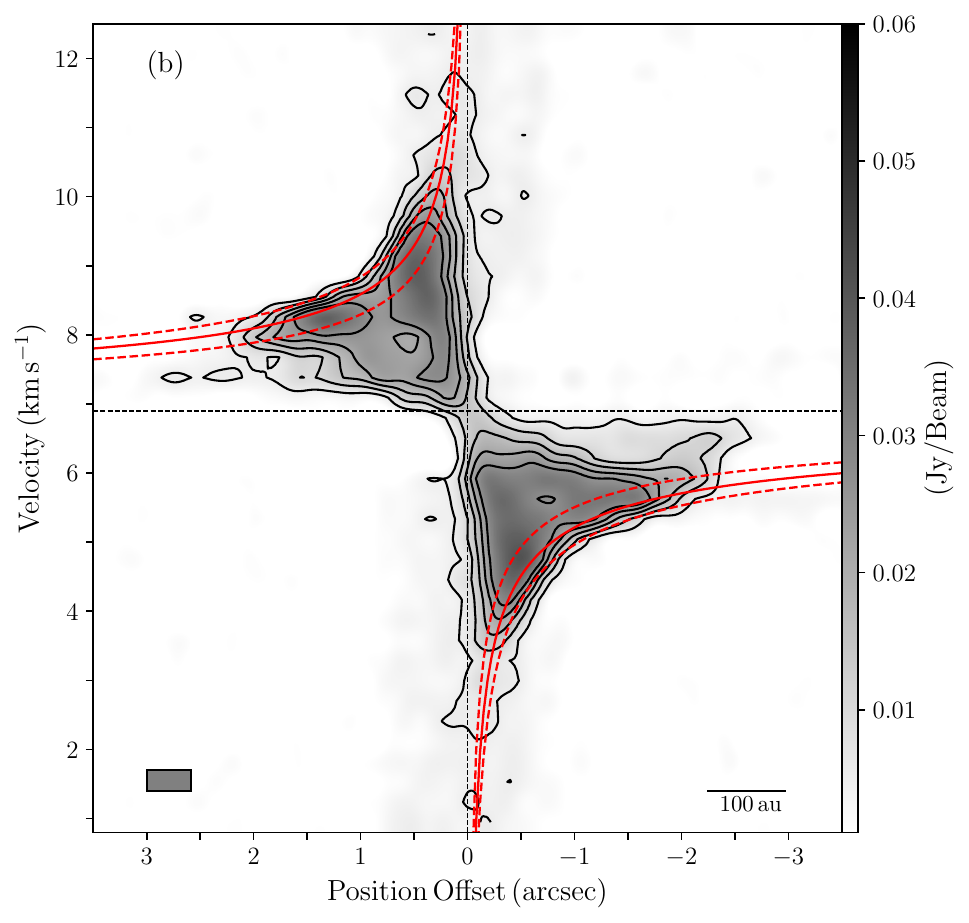}
\caption{HH 30 disk emission of $^{13}$CO ($J=$2--1) molecular line. (a) ALMA first moment or the intensity-weighted velocity of the accretion disk overlaid by contours of the moment zero (integrated intensity). Contours levels start from 5$\sigma$ in steps of 5$\sigma$, 10$\sigma$, 15$\sigma$, and 20$\sigma$, where $\sigma=2.13\times10^{-3}$ Jy/Beam $\kms$. The synthesized beam of the disk image is shown in the lower left corner of the panel. The moments was integrated in a range of the velocities from 3 $\kms$ to 11 $\kms$. (b) Position--velocity diagram over the disk mid-plane (horizontal black dashed line in panel a). The red solid lines show a Keplerian velocity profile surrounding the 0.45 $\msun$ central object, while the red dashed lines represent a Keplerian velocity profiles corresponding to $0.31\,\msun$ and $0.59\,\msun$ (inner/outer curves), respectively. The gray bar represents the angular resolution (0.3$^{\prime \prime}$ or 42 $\au$) and the channel width (0.3 $\kms$). Contour levels start from 5$\sigma$ in steps of 5$\sigma$, 10$\sigma$, 15$\sigma$, and 20$\sigma$, where $\sigma=1.07\times10^{-3}$ Jy/Beam. }
\label{fig:pvdisk}
\end{figure*}

The ratio between the mass-loss rate and the accretion rate is thought to be around $\sim$10\% (\citealt{Ellerbroek2013}), although, recent studies have shown that this ratio can be around $\sim$50\% (\citealt{Lee2020}; \citealt{Podio2021}). However, for many sources a constant accretion rate implies that the source does not reach the final mass consistent with the initial mass function (\citealt{Evans2009}; \citealt{Caratti2012}). This scenario may suggest that the accretion may be episodic (\citealt{Frank2014}).
The greatest evidence of the episodic ejections from the accretion disk has been observed in the jets that show a series of knots along their axes, such as, HH 211 (\citealt{Lee2007}), CARMA-7 (\citealt{Plunkett2015}), and HH 212 (\citealt{Lee2017}). Through recent high-resolution ALMA observations, the evidence of the episodic ejections has been reported by the molecular outflows HH 46/47 (\citealt{Zhang2019}), DO Tauri (\citealt{FernandezLopez2020}), and DG Tau B (\citealt{deValon2022}), in which the authors show that the molecular outflow has an internal multiple shell structure, where the outer shell could be associated with the swept-up material by the disk-wind, while the internal shells are associated with short episodic wind or jet outburst ejected directly from the accretion disk every few 100 yr. 

Located in the dark cloud L1551 at a distance of $\sim$141 $\pm$7 pc \citep{Zucker2019} in Taurus, the HH 30 is a young molecular outflow associated with a T Tauri star with an enclosed mass of 0.45$\pm$0.04 M$_\odot$ and a spectral class M0$\pm$1 (\citealt{Pety2006}), the central source has a bolometric luminosity of 0.2--0.9 $\lsun$ (\citealt{Cotera2001}), and the velocity $V_\mathrm{lsr}=6.9\pm0.1\,\kms$ \citep{Louvet2018}. The HH 30 system is a typical protostellar object with an edge-on accretion disk with a mass of dust of $>25.5$ M$_\oplus$ (\citealt{Villenave2020}). The ballistic jet of the HH 30 has a size of $7^\prime$ and presents wiggling and orbital motions of the central star in a binary system with a period $<1$ yr (\citealt{Anglada2007}), and a large-scale C-shape due to proper motion of the system or due to the action of winds from other stars (\citealt{Estalella2012}). Whereas the optical jet is bipolar, the outflow is only seen north of the disk, constituting an example of a monopolar outflow (e.g., \citealt{FernandezLopez2013}), possibly due to the lack of molecular material south of the disk, that is, the HH 30 may be located at the boundary of the parental core (\citealt{Stanke2022}). In previous work, \citet{Louvet2018} focus on studying a region of the HH 30 outflow near to central protostar (i.e., at heights lower than $z\sim3.0^{\prime \prime}$). They found that the properties of the outflow can be explained by a slow-disk wind.

Here we present ALMA observations of the emission of the molecular lines of $^{12}$CO (J=2--1) and $^{13}$CO (J=2--1) of the protostellar outflow and the accretion disk associated with HH 30. In this work, we analyze the morphology and the kinematics of the outflow at larger scales. The paper is organized as follows. Section \ref{sec:observations} details the observations. The results are shown in Section \ref{sec:results}. In Section \ref{sec:discussion}, we discuss our results. Finally, the conclusions are presented in Section \ref{sec:conclusions}.

\section{Observations} \label{sec:observations}

The archival observations of HH 30 were carried out with the Atacama Large Millimeter/Submillimeter Array (ALMA) in band 6 in 2015 July on 19 and 21 as part of the program 2013.1.01175.S (P.I. Catherine Dougados) and in band 6 in 2018 on October 21 and 22, on November 07 and 10, as part of the program 2018.1.01532.S (P.I. Fabien Louvet) at the phase centre $\alpha$(J2000)=$04^\mathrm{h}31^\mathrm{m}37^\mathrm{s}.47$ and $\delta$(J2000)=$+18^{\circ}12^{\prime}24^{\prime\prime}.22$.

The integration time on-source was about 106 minutes, and 34 minutes was used for calibration for the 2015 observations, while for the 2018 observations it was about 106 minutes on-source and 212 minutes for calibrations. 
For the 2015 observations, the ALMA digital correlator was configured with five spectral windows centered on 230.546 GHz (spw0), 234 GHz (spw1), 220.379 GHz (spw2), 219.562 (spw3), and 217.052 GHz (spw4), with 960 channels and a space channel of 122 kHz or about 0.17 km s$^{-1}$ for spw0, spw2, and spw3, and with 128 channels and a space channel of 15.625 MHz or about 21.5 $\kms$ for the continuum spectral windows (spw1 and spw4). For the observations of 2018 the correlator was configured with six spectral windows centered on 233.994 GHz (spw0), 231.214 GHz (spw1), 230.531 GHz (spw2), 216.994 GHz (spw3), 220.393 GHz (spw4), and 219.554 GHz (spw5). The spectral windows spw1, spw2, spw4, and spw 5 have 480 channels of 122 kHz or about 0.17 km s$^{-1}$, while the continuum spectral windows (spw 0 and spw3), have 64 channels of 31.250 MHz width or about 40.19 $\kms$ and 960 channels of 1.953 MHz or 2.69 $\kms$. The spectral lines reported on this study were found in spw0 ($^{12}$CO) and spw2 ($^{13}$CO) for 2015 observations and spw2 ($^{12}$CO) and spw4 ($^{13}$CO) for 2018 observations.

For both observations, the weather conditions were reasonably good and stable with a mean value PWV$\approx$0.9 mm for these high frequencies. The observations used the quasars J0423--0120, J0423--013, J0522--3627, J0510+1800, and J0502+1338 for amplitude, phase, bandpass, pointing, water vapor radiometer, and atmosphere calibration.

\begin{figure*}[t!]
\centering
\includegraphics[scale=0.585]{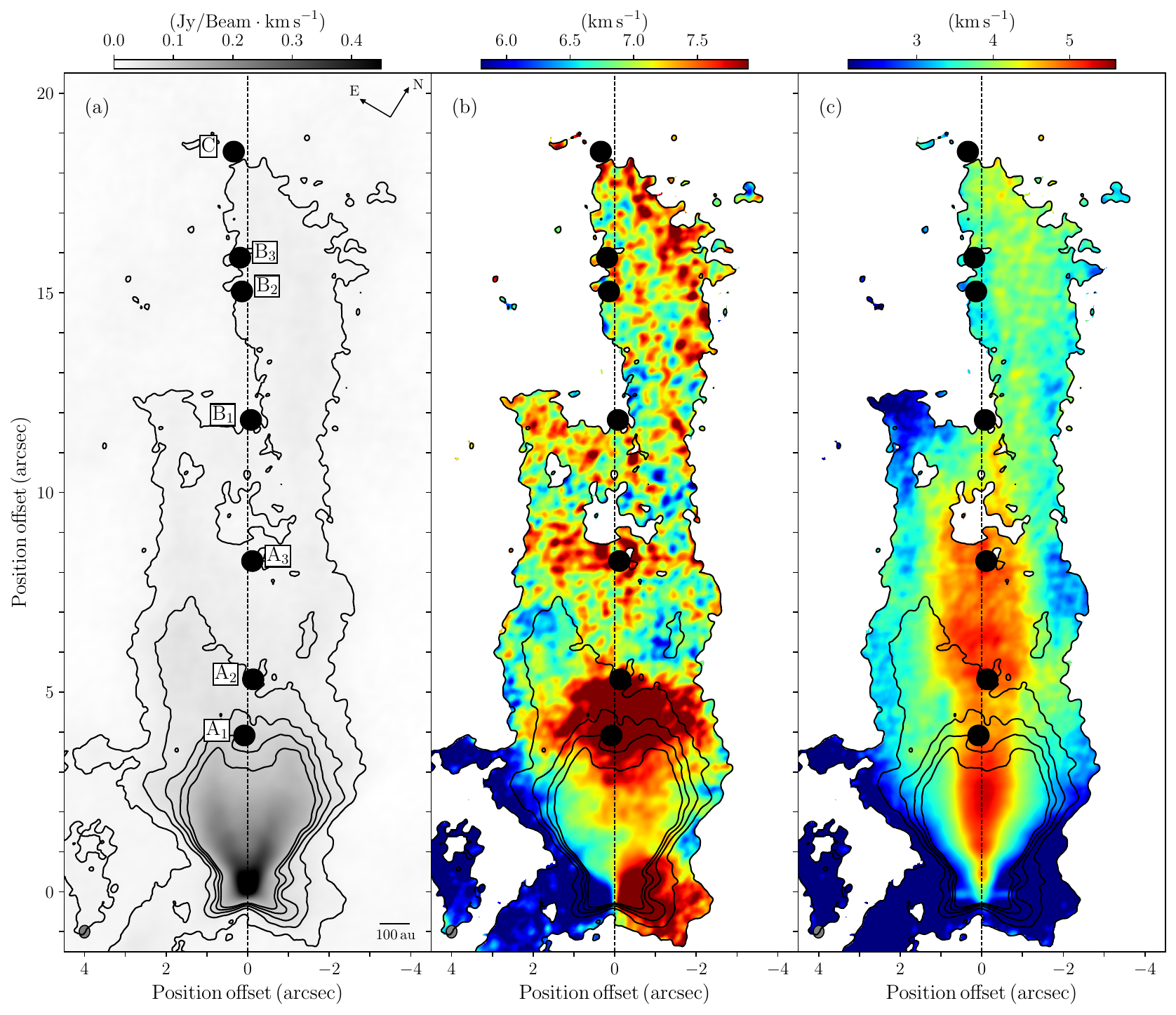}
\caption{HH 30 molecular line emission of $^{12}$CO ($J$=2--1) molecular line. (a) ALMA moment zero or integrated intensity. (b) ALMA first moment or the intensity-weighted velocity. (c) ALMA second moment or the intensity-weighted dispersion velocity. The black dots represent the [S$_\mathrm{II}$] jet knots reported by \citet{Anglada2007} and \citet{Estalella2012}. The knots in all panels are corrected by the proper motions. The synthesized beam is shown in the lower left corner. The contour levels in the three panels start from 3$\sigma$ in steps of 3$\sigma$, 6$\sigma$, 9$\sigma$, and 12$\sigma$, where $\sigma$=6.69$\times$10$^{-3}$ Jy/Beam $\kms$.}
\label{fig:knots_moments}
\end{figure*}

The data were calibrated using the common astronomy software application (CASA) package (\citealt{CASAteam})  version 4.3.1 and version 5.4.0 for the project 2013.1.01175.S and 2018.1.01532.S, respectively. After calibration the data were selfcalibrate. We combined the data from the two observation programs and produced images using a robust parameter of 0.5. The final velocity cubes have a rms noise level of 1.07 mJy/Beam and 0.96 mJy/Beam for $^{13}$CO and $^{12}$CO data respectively. Finally the angular resolution is 0.32$^{\prime \prime}$ $\times$ 0.26$^{\prime \prime}$ with a PA of -6.91$^{\circ}$ and 0.32$^{\prime \prime}$ $\times$ 0.27$^{\prime \prime}$ with a PA of -5.93$^{\circ}$, for the $^{13}$CO and the $^{12}$CO cubes, respectively.

\section{Results} \label{sec:results}
\subsection{Disk emission}
\label{subsec:diskemission}

Figure \ref{fig:pvdisk} shows  $^{13}$CO molecular line emission of the disk associated with the HH 30 system. The color map of Figure \ref{fig:pvdisk}a presents the first moment or the intensity-weighted velocity map overlaid in black contours by the moment zero map. The east side of the disk has blueshifted velocities, while the west side presents redshifted velocities. This difference in the velocities is evidence of the rotation of the disk.

Figure \ref{fig:pvdisk}b is the position-velocity diagram along the disk mid-plane. The fits correspond to Keplerian curves $v_k=\sqrt{G M_{\mathrm{dyn}}/r}$, where $G$ is the gravitational constant and $r$ is the radius. In order to get the best fit, we used the pvanalysis package of the Spectral Line Analysis/Modeling (SLAM) code (\citealt{Aso2023}). The pvanalysis tool extracts rotation curves based on the methods using edge (\citealt{Seifried2016}) and ridge (e.g., \citealt{Aso2015}; \citealt{Sai2020}) of the emission in the position-velocity diagram.
%
The red solid lines correspond to a dynamical mass of $M_\mathrm{dyn}=0.45\pm0.14\,\msun$. This value is the best fit of the average between the dynamical masses obtained using the peaks of the emission $M_\mathrm{dyn}=0.31\,\msun$ (inner dashed lines) and the 5$\sigma$ limit of the emission $M_\mathrm{dyn}=0.59\,\msun$ (outer dashed lines), the dynamical mass obtained with the average of the best fits matches the value previously reported by \citet{Pety2006}.

\subsection{Molecular outflow emission}
\label{subsec:outflowemission}

\begin{figure*}[t!]
\centering
\includegraphics[scale=0.3]{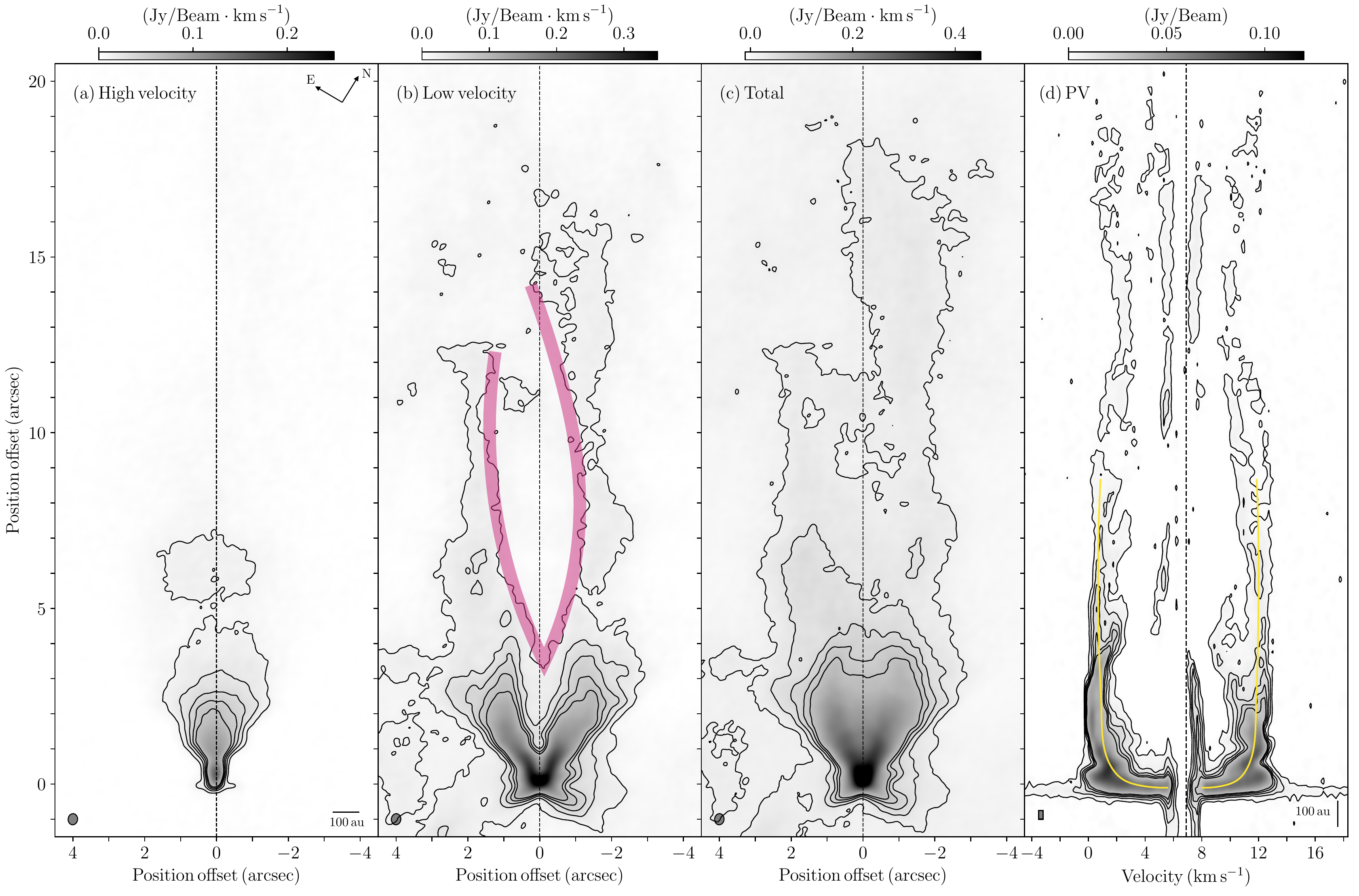}
\caption{ALMA moment zero of the molecular outflow associated with HH 30 integrated at difference ranges of the line-of-sight velocity. (a) High-velocities, from -0.1 $\kms$ to 2.3 $\kms$ and from 11.3 $\kms$ to 14 $\kms$. (b) Low-velocities, from 2.6 $\kms$ to 11 $\kms$. (c) Full range of the emission, from -0.1 $\kms$ to 14 $\kms$. (d) Position-velocity diagram along the jet axis. The synthesized beam of panels (a), (b), and (c) is shown in the lower left corner. The magenta line in panel (b) traces an internal cavity of the molecular outflow. The yellow lines in panel (d) indicate the convex spur structure and the gray bar represents the angular resolution (0.3$^{\prime \prime}$ or 42 $\au$) and the channel width (0.3 $\kms$) used for the position–velocity cut. The contour levels in the panels (a)--(c) start from 3$\sigma$ with steps of 3$\sigma$, 6$\sigma$, 9$\sigma$, and 12$\sigma$, where $\sigma$=3.69$\times$10$^{-3}$ Jy/Beam $\kms$ (panel a), $\sigma$=3.93$\times$10$^{-3}$ Jy/Beam $\kms$ (panel b), $\sigma$=5.87$\times$10$^{-3}$ Jy/Beam $\kms$ (panel c). The contour levels in panel (d) start at 3$\sigma$ with steps of 6$\sigma$, 12$\sigma$, 18$\sigma$, and 24$\sigma$, where $\sigma$=9.62$\times$10$^{-4}$ Jy/Beam.}
\label{fig:mom0_comp_LHVC}
\end{figure*}

The emission of the $^{12}$CO of the molecular outflow associated with HH 30 is shown in Figure \ref{fig:knots_moments}, the images of this figure are rotated by an angle of 31.6$^{\circ}$, the position of the jet axis (\citealt{Anglada2007}). Figure \ref{fig:knots_moments}a shows the ALMA moment zero, the emission of the outflow extends up to $\sim$19$^{\prime\prime}$.
The black dots denoted by A$_{1,2,3}$, B$_{1,2,3}$, and C correspond to the [S$_\mathrm{II}$] jets knots reported by \citet{Anglada2007}, the positions of the jet knots are corrected by the proper motions measured by \citet{Estalella2012} and considering a position offset corresponding to 8 yr, which is the difference between their observations made on 2010 and the observations reported in this work made on 2018. The ALMA first moment of the outflow is presented in Figure \ref{fig:knots_moments}b where the molecular outflow presents signatures consistent with rotation of the gas at heights $z\leq$ 4$^{\prime\prime}$, and the rotation velocity is $\lesssim$0.5 $\kms$. In this region, the outflow has blueshifted velocities on the east side and redshifted velocities on the west side. 
For very low heights $z<$1$^{\prime \prime}$, the rotation of the gas is dominated by the accretion disk. The high velocity observed ($> 8\,\kms$) at a height 4$^{\prime\prime}<z<$6$^{\prime \prime}$ of the gas, between the knots A$_1$ and A$_2$, could be associated with entrained material from these knots. Figure \ref{fig:knots_moments}c shows the dispersion velocity or moment two map. The inner part of the outflow presents larger velocity dispersion than the walls, this effect may be due to the interaction of the innermost high-velocity jet (the jet seen at optical wavelengths) with the molecular environment or the material from wide-angle wind. 
The fact that in panel (c) we do not observe the high velocity presented in panel (b) at a height $4^{\prime \prime}<z<6^{\prime\prime}$ could be due to the dispersion velocity in this region is similar to the dispersion velocity caused by the high-velocity jet.

\begin{figure*}[t!]
\centering
\includegraphics[scale=0.438]{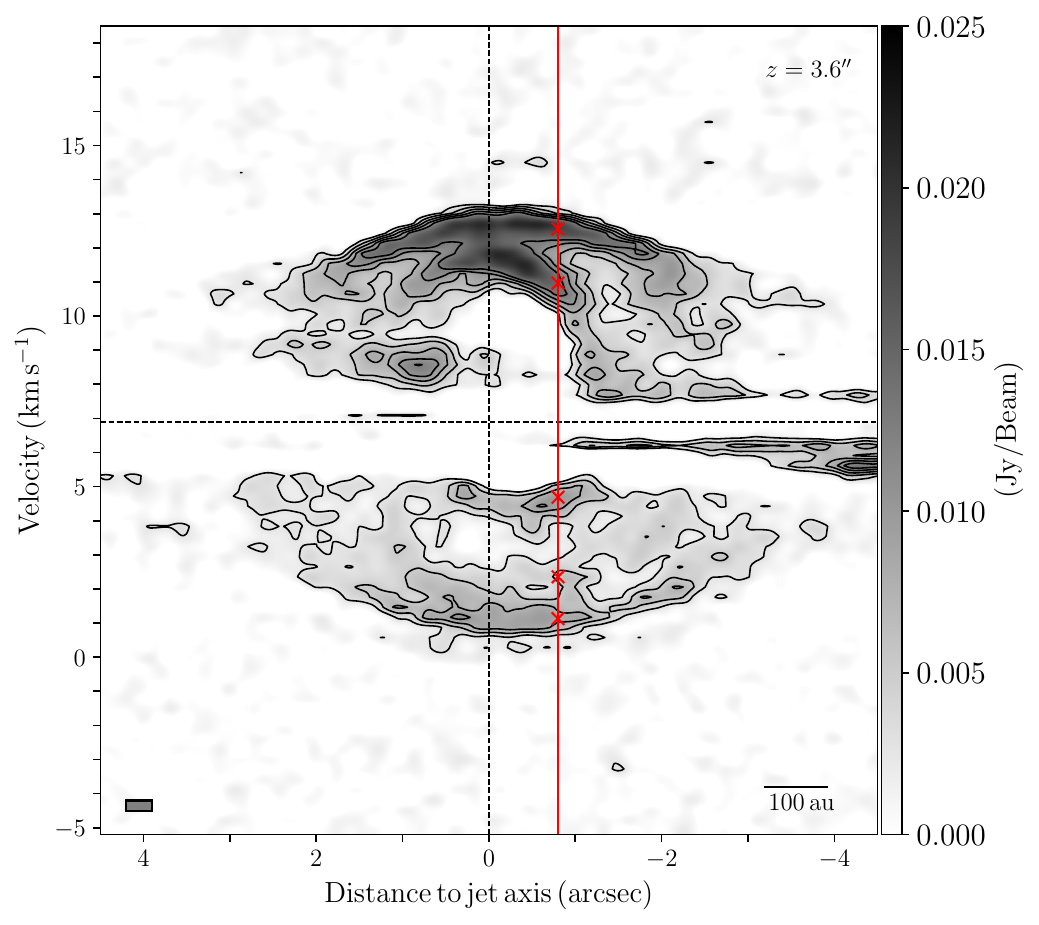}\includegraphics[scale=0.438]{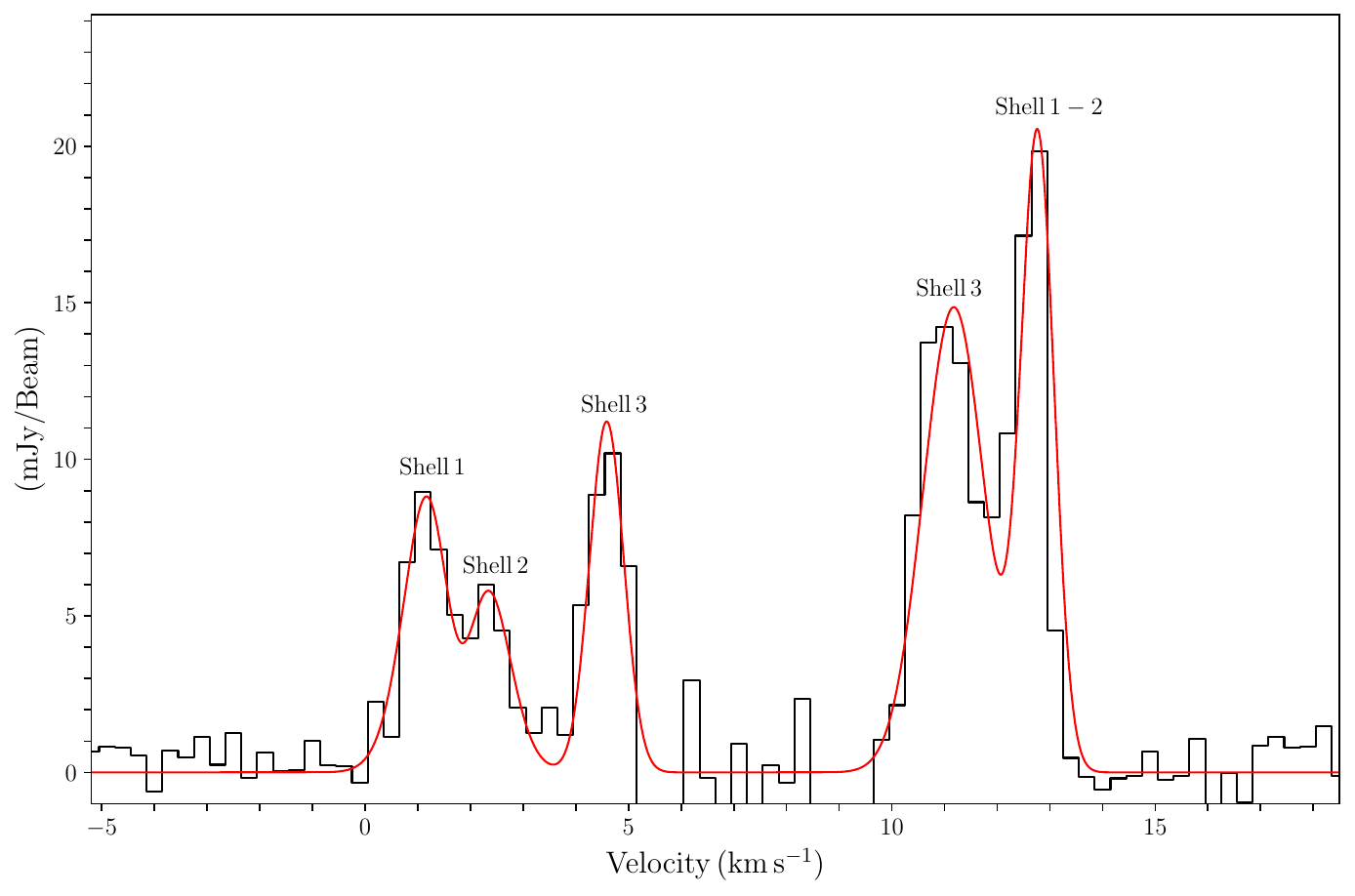}
\caption{\textit{Left panel}: Position-velocity diagram perpendicular to the jet axis at a height of $z=3.6^{\prime \prime}$ above the disk mid-plane. The contours start at 3$\sigma$ in steps of 3$\sigma$, 6$\sigma$, 9$\sigma$, and 12$\sigma$, where $\sigma$=0.96$\times$10$^{-3}$ Jy/Beam. The vertical dashed line represents the position of the jet axis, while the horizontal line is the $V_\mathrm{lsr}$=6.9 $\kms$. The gray bar represents the angular resolution (0.3$^{\prime \prime}$ or 42 $\au$) and the channel width (0.3 $\kms$). The red line denotes a vertical cut along which the spectrum of the right panel was obtained and the red crosses are the position of the peaks of the emission. \textit{Right panel}: Spectrum obtained through vertical cut of the position-velocity diagram of left panel. The black solid line represents the emission observed and the red solid line is the best five-Gaussian fits.}
\label{fig:cuts}
\end{figure*}

Figure \ref{fig:mom0_comp_LHVC} shows the ALMA moment zero of the molecular outflow integrated at different velocity regimes. Figure \ref{fig:mom0_comp_LHVC}a shows the moment zero integrated from -0.1 $\kms$ to 2.3 $\kms$ and from 11.3 $\kms$ to 14 $\kms$. These ranges correspond to high velocities with respect to the velocity $V_\mathrm{lsr}=$ 6.9 $\pm$ 0.1 $\kms$. This Figure shows the emission of the outflow close to the accretion disk, while the walls of the cavity outflow (magenta line) are shown in Figure \ref{fig:mom0_comp_LHVC}b, where the moment zero is integrated at velocities close to the $V_\mathrm{lsr}$ velocity, from 2.6 $\kms$ to 11 $\kms$. The full range emission is presented in Figure \ref{fig:mom0_comp_LHVC}c. Finally, Figure \ref{fig:mom0_comp_LHVC}d is the position-velocity diagram along the jet axis. This diagram shows a convex spur structure which is the signature of the jet-driven bow shocks (\citealt{Lee2001})
These bow shocks could be associated with the S$_\mathrm{[II]}$ knots of the Figure \ref{fig:knots_moments}, however, they are not detected with our observations. The position-velocity diagram may trace the walls of the molecular outflow which tend to have a constant velocity, also, this diagram shows a possible internal structure at a height $2 ^{\prime \prime}\lesssim z\lesssim 5^{\prime \prime}$, the internal structure could be associated with gas at different velocities ($8\,\kms \lesssim V \lesssim 14\,\kms$). The apparent lack of the emission in the surroundings of the $V_\mathrm{lsr}$ velocity could be associated with the absorption by sightly colder component at $V=V_\mathrm{lsr}$ in front of HH 30.

In addition, position-velocity diagrams perpendicular to the jet axis at different heights above the disk mid-plane were made.
The left panel of Figure \ref{fig:cuts} show, as an example, a position-velocity diagram at a height of $z$=3.6$^{\prime\prime}$. The black dashed lines represent the jet axis (vertical line) and the velocity $V_\mathrm{lsr}$ (horizontal line). The solid red line is a vertical cut made to obtain the spectrum shown in the right panel of Figure \ref{fig:cuts}. 
In the spectrum, five peaks can be observed, 
that may be correlated with the emission of three putative different shells. This cut is shown as an example to explain our method to detect these shells. To ensure that the outflow has internal multiple shells, 
we selected different spectra as the one shown in the right panel of Figure \ref{fig:cuts}.

\begin{figure*}[t!]
\centering
\includegraphics[scale=0.485]{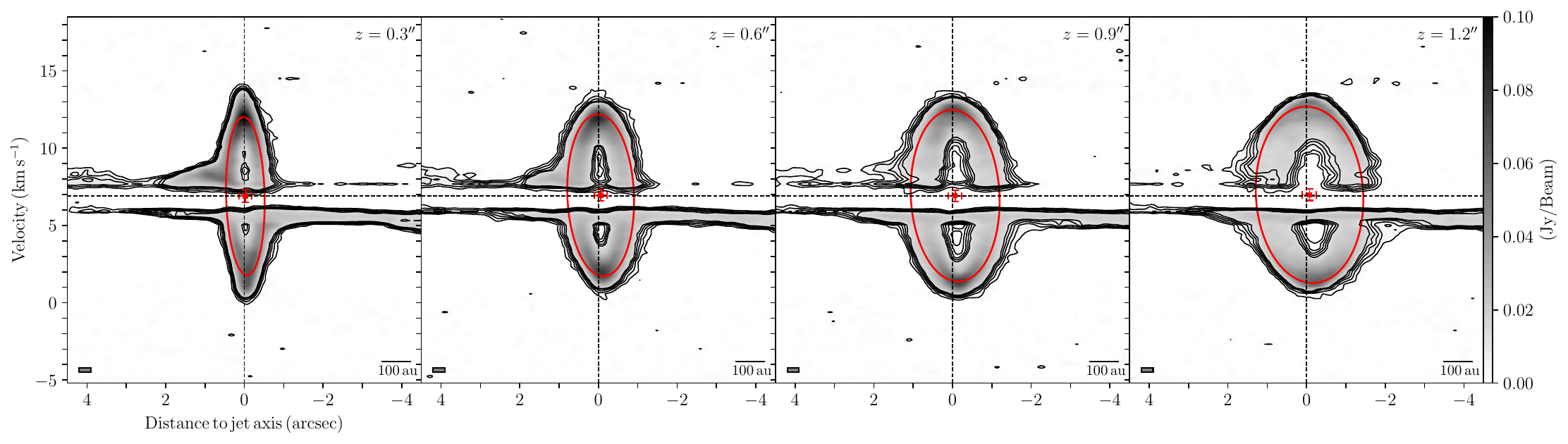}
\includegraphics[scale=0.485]{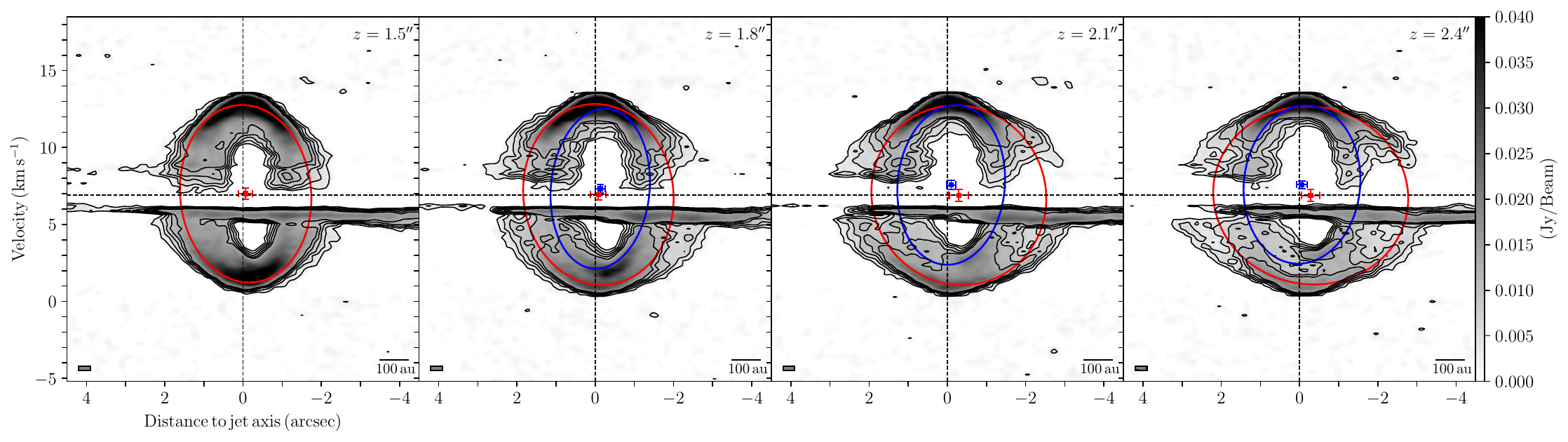}
\includegraphics[scale=0.485]{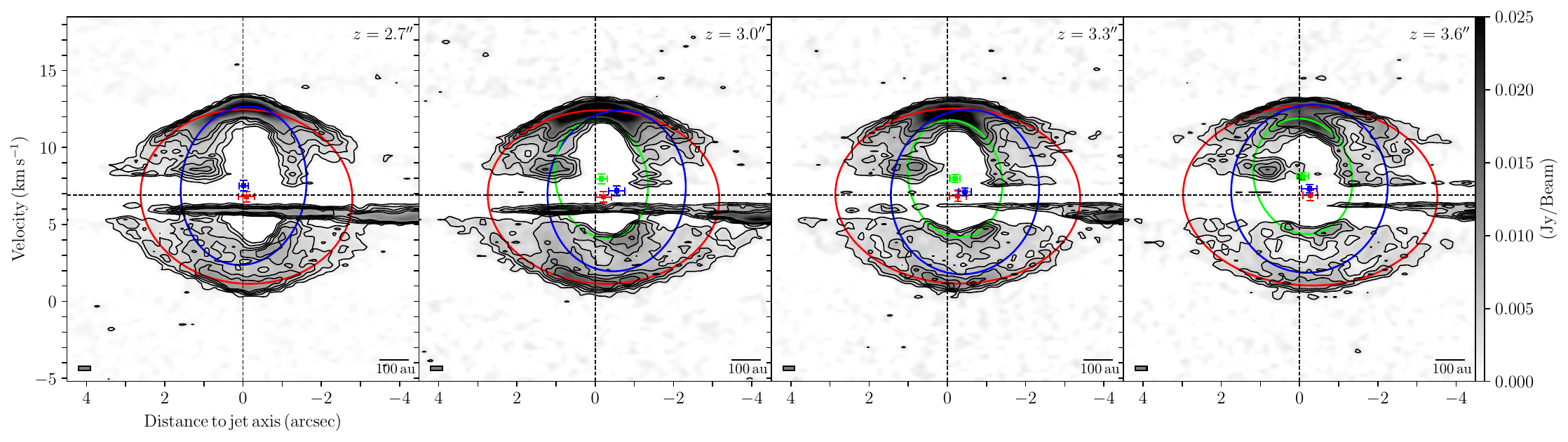}
\includegraphics[scale=0.485]{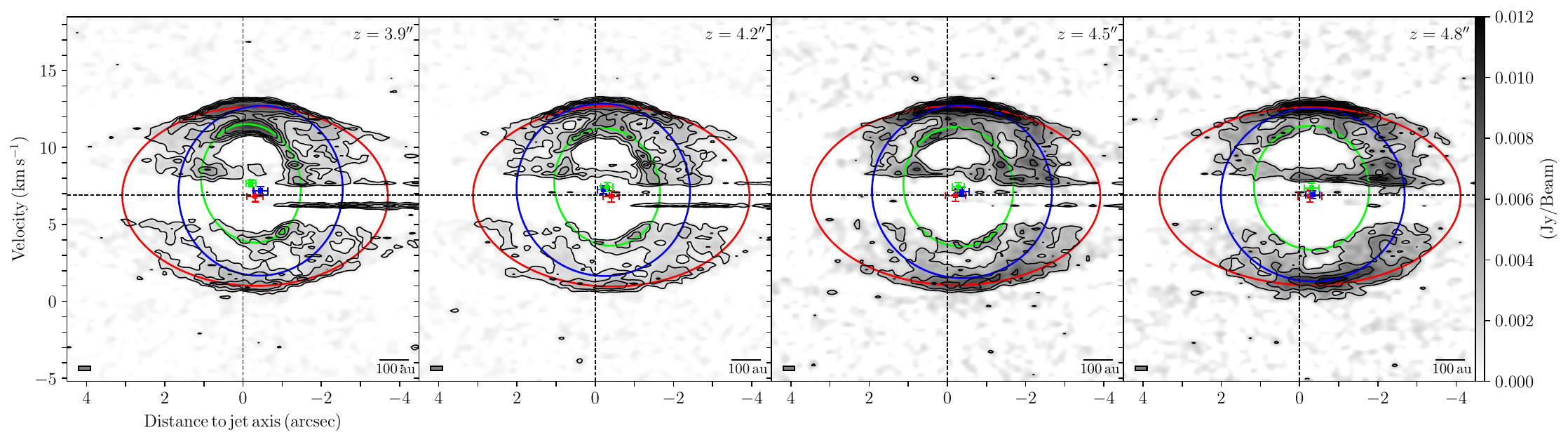}
\caption{Position-velocity diagrams from the $^{12}$CO emission perpendicular to jet axis at different heights from $z$=0.3$^{\prime\prime}$ (42 $\au$) to 4.8$^{\prime\prime}$ (672 $\au$) with an interval of 0.3$^{\prime \prime}$ (42 $\au$). The gray bar represents the angular resolution (0.3$^{\prime \prime}$ or 42 $\au$) and the channel width (0.3 $\kms$). The ellipses in the different panels represent the best fit for the shell 1 (red), shell 2 (blue), and shell 3 (green), and the crosses show the center of these ellipses. The contours levels start at 3$\sigma$ in steps of 3$\sigma$, 6$\sigma$, 9$\sigma$, and 12$\sigma$, where $\sigma$=0.96$\times$10$^{-3}$ Jy/Beam. The vertical dashed line represents the position of the jet axis, while the horizontal line is the $V_\mathrm{lsr}$=6.9 $\kms$.}
\label{fig:pvmoments}
\end{figure*}

Position-velocity diagrams perpendicular to the jet axis at different heights are shown in Figure \ref{fig:pvmoments}. The diagrams were made from $z=0.3^{\prime\prime}$ (or $\sim$42 $\au$) to $z=4.8^{\prime \prime}$ (or $\sim$672 $\au$) every 0.3$^{\prime\prime}$. At heights close to the accretion disk ($z\leq 1.5^{\prime \prime}$), we only detect one shell, while for intermediate heights ($1.8^{\prime \prime}\leq z \leq 2.7^{\prime \prime}$), we can distinguish two shells. Finally, for high-heights ($z\geq3.0^{\prime\prime}$), we detect three possible shells. 
In a previous work, \citet{Louvet2018} found an inner shell at a height of $z=2.25^{\prime \prime}$, with the observations reported in this work, we confirmed the presence of this shell.
Such shell structure is consistent with radially expanding shells or bubbles (\citealt{Arce2011}; \citealt{Zapata2011}; \citealt{Zapata2014}) and is similar to the elliptical structure expected in position-velocity diagrams of an outflow with a low inclination angle respect to the plane of the sky (\citealt{Lee2000}).
We made cuts in each position-velocity diagrams at different positions every 0.15$^{\prime\prime}$ and extract the spectra from each of these cuts. We did a Gaussian fit to the spectra and identified peaks and correlate them with a structure in position, defining a new point in the position-velocity diagram with its error bar. We fit ellipses to the points associated with the peaks of the spectra assuming that these points trace a single structure. The fitted ellipses are obtained using lsq-ellipse package from python. We have named these shells as 1, 2, and 3 for the red, blue, and green ellipses, respectively.
As mentioned above, the elliptical shape of the three shells is evidence that the three shells are in radial expansion and the expansion velocity does not vary with distance to the protostar. The inclination of the shells with respect to jet axis (vertical dashed lines) is an evidence of the rotation. The signatures of the rotation is more evident in the shells 2 and 3.   

\subsection{Kinematic and physical properties of the molecular outflow}\label{subsec:kinematicoutflow}

\begin{figure*}[t!]
\centering
\includegraphics[scale=0.5]{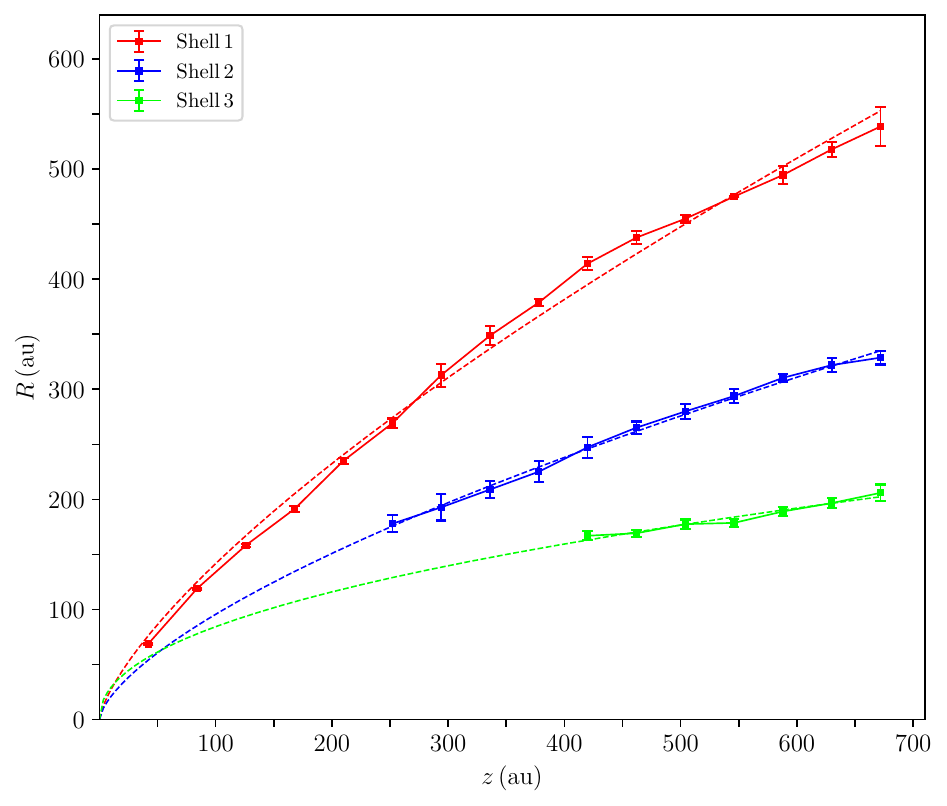}\includegraphics[scale=0.5]{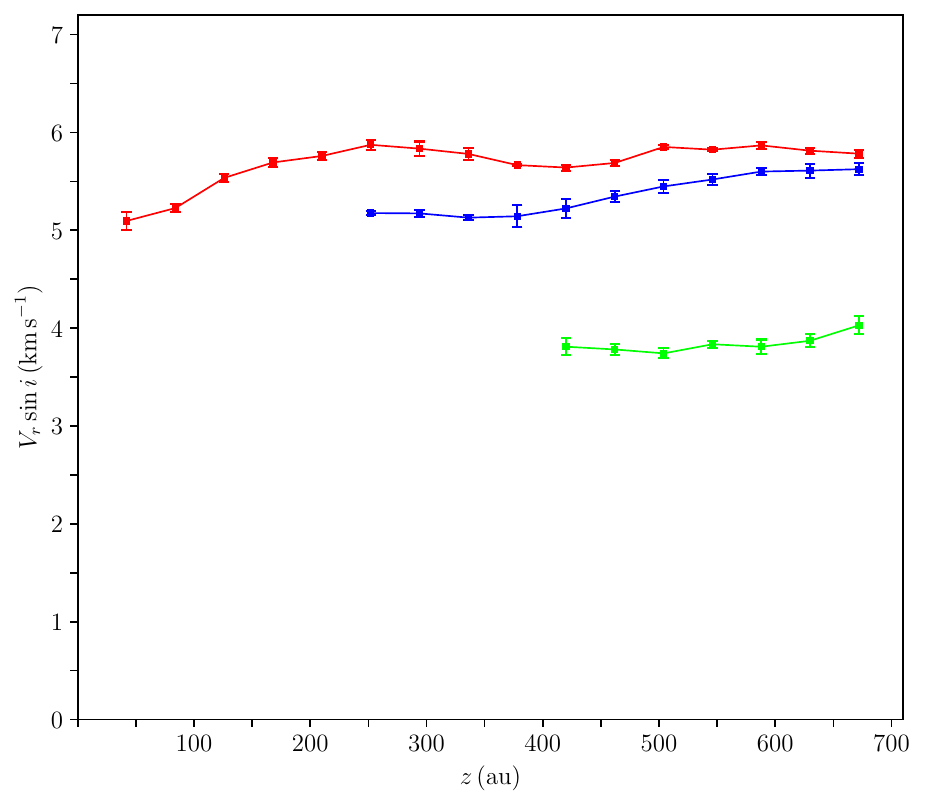}
\includegraphics[scale=0.5]{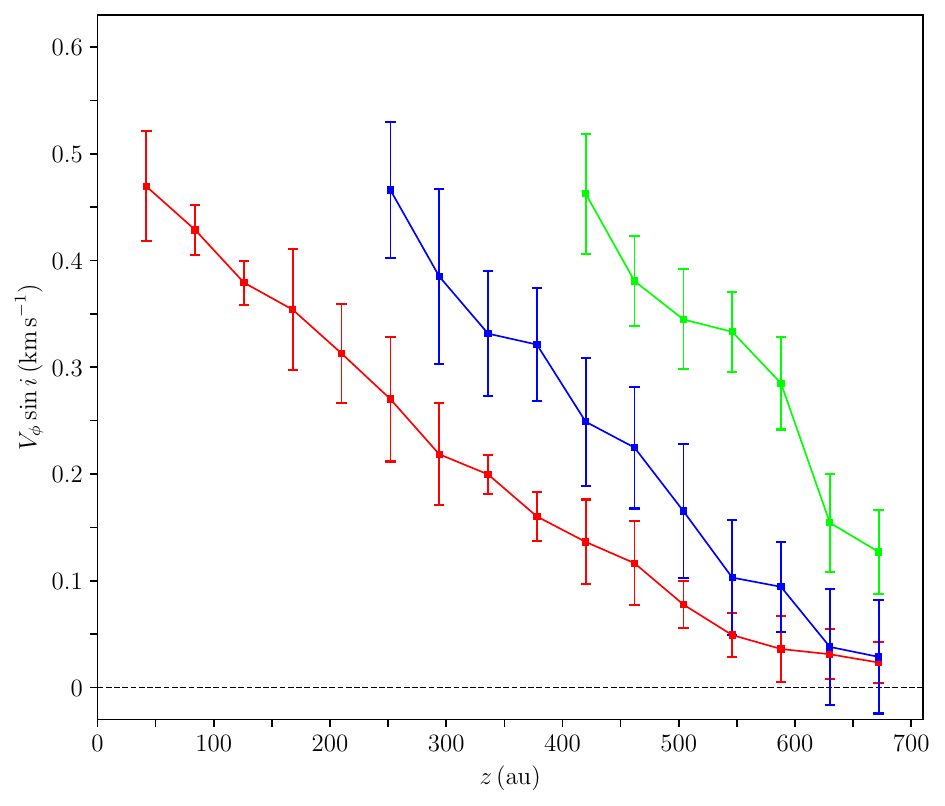}\includegraphics[scale=0.5]{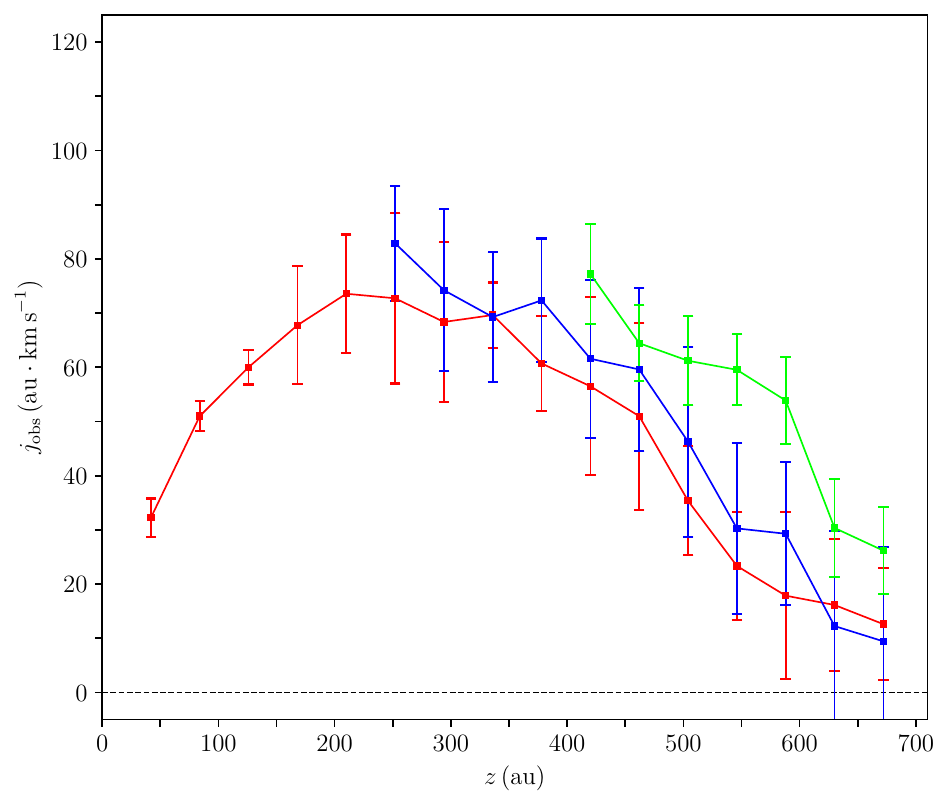}
\caption{The kinematical properties of the HH 30 molecular outflow for the different shells as a function of the height $z$. \textit{Top left panel}: the cylindrical radius. \textit{Top right panel}: the expansion velocity perpendicular to the jet axis. \textit{Bottom left panel}: the rotation velocity. \textit{Bottom right panel}: the specific angular momentum.
The error bars are derived from the Gaussian fit (see text). The dotted lines in the top left panel correspond to the best fitting of the general relation $z=a R^{-\beta/2}$ where the values of $a$ and $\beta$ for each shell are shown in the text.}
\label{fig:parameters}
\end{figure*}

\begin{figure*}[t!]
\centering
\includegraphics[scale=0.32]{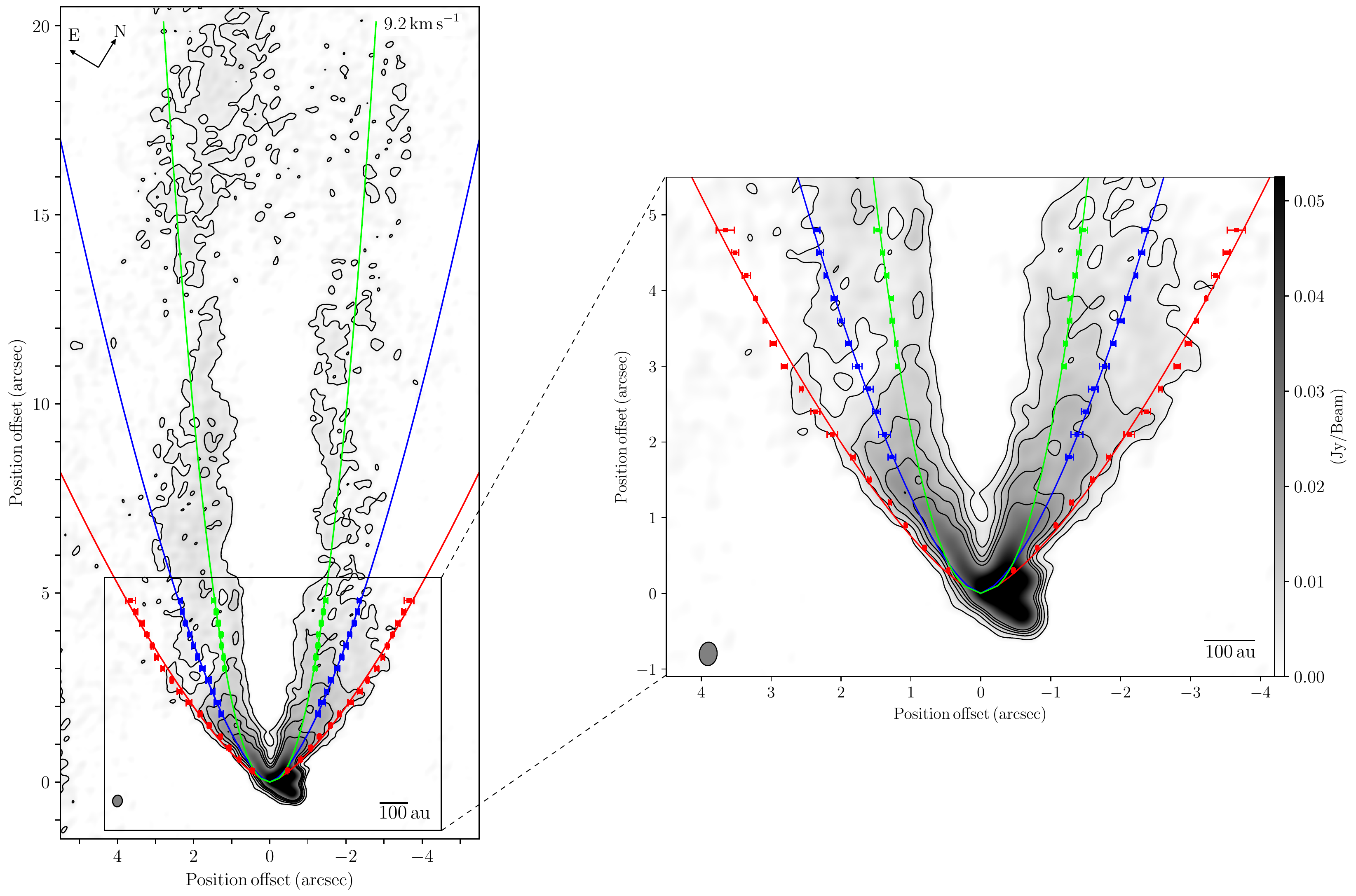}
\caption{The $^{12}$CO molecular line emission of the velocity channel of 9.2 $\kms$ of the molecular outflow associated with HH 30. \textit{Left panel:} emission of all the detected outflow emission. \textit{Right panel:} zoom into the central region. The solid lines and points show the position of the different shells, shell 1 (red lines), shell 2 (blue lines), and shell 3 (green lines). The synthesized beam in all panels is shown in the lower left corner. The contours levels start at 3$\sigma$ in steps of 3$\sigma$, 6$\sigma$, 9$\sigma$, 12$\sigma$, and 15$\sigma$, where $\sigma$=0.96$\times$10$^{-3}$ Jy/Beam.}
\label{fig:channels}
\end{figure*}

To determine the kinematic and physical properties of the molecular outflow, we use the outflow model presented by \citet{Louvet2018} (hereafter Louvet's model). In this model, they consider that for each height $z$ the $^{12}$CO emission arises from a narrow circular ring of gas. Louvet's model relates the physical properties of the circular ring, such as the radius $R$, the center $x_\mathrm{offset}$, and the velocity field ($V_r$, $V_z$, and $V_\phi$), with the parameters of an ellipse, position angle (PA), major $a$ and minor $b$ axes, and the center coordinates ($r_\mathrm{cent}$ and $V_\mathrm{cent}$). This model considers the wiggling movements and the inclination angle with respect to the line of sight $i$. The equations of Louvet's model are

\begin{eqnarray}
    x_\mathrm{offset}=r_\mathrm{cent},
    \label{eq:xoffset}
\end{eqnarray}
\begin{eqnarray}
    V_z=-\left(V_\mathrm{cent}-V_0\right)/\cos i,
    \label{eq:vz}
\end{eqnarray}
\begin{eqnarray}
    \left(V_r\sin i\right)^2=\left(\left(\cos\mathrm{PA}\right)^2/a^2 +\left(\sin\mathrm{PA}\right)^2/b^2\right)^{-1},
    \label{eq:vrsini}
\end{eqnarray}
\begin{eqnarray}
    \left[\left(V_\phi\sin i\right)/R\right]&=&0.5\times\left(V_r\sin i\right)^2\times\sin\mathrm{2PA}\nonumber \\
    &\times&\left(1/b^2-1/a^2\right),
    \label{vphisini}
\end{eqnarray}
\begin{eqnarray}
    \left(1/R^2\right)&=&\left(\left(\cos\mathrm{PA}\right)^2/b^2+\left(\sin\mathrm{PA}\right)/a^2\right)\nonumber \\
    &-&\left(V_\phi/R\right)^2/V_r^2,
    \label{eq:rcyl}
\end{eqnarray}
where we can consider that $x_\mathrm{offset}$ is the distance between the horizontal center coordinate of the ellipse and the jet axis, and $V_\mathrm{cent}$ is the velocity offset between the vertical center coordinate and the $V_\mathrm{lsr}$ velocity. Finally, $V_0$ is the projected cut velocity along the line of sight. In this case, $V_0=0$ $\kms$ because we assume that the outflow is not wiggling.

\begin{figure*}[t!]
\centering
\includegraphics[scale=0.54]{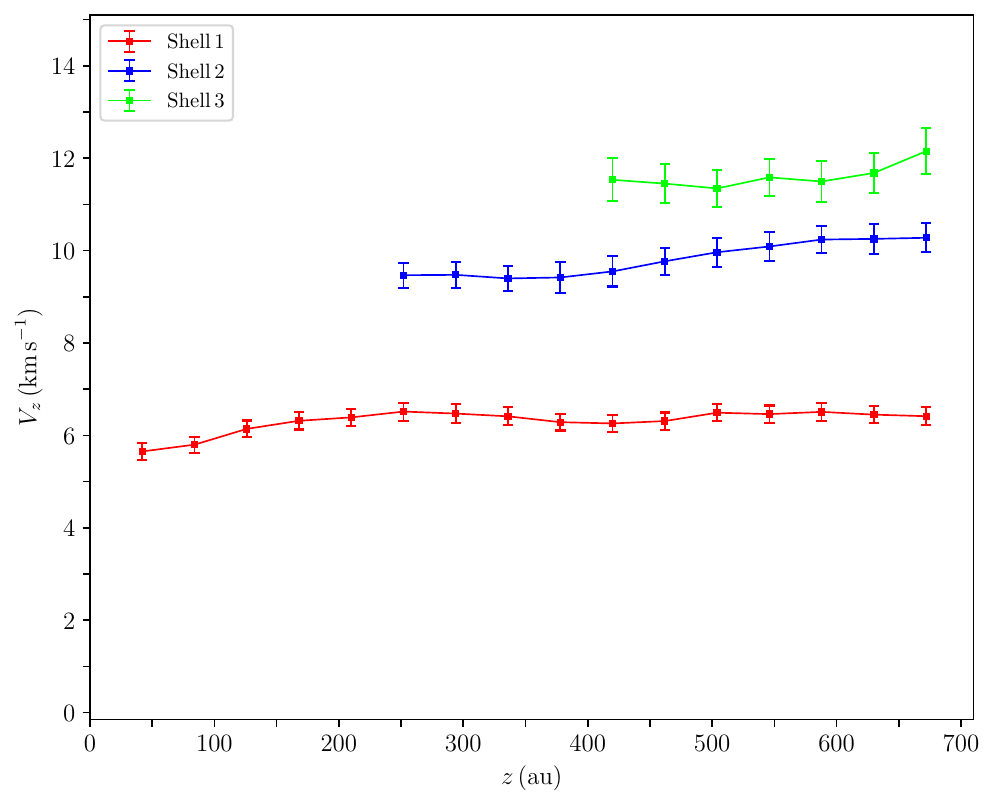}\includegraphics[scale=0.54]{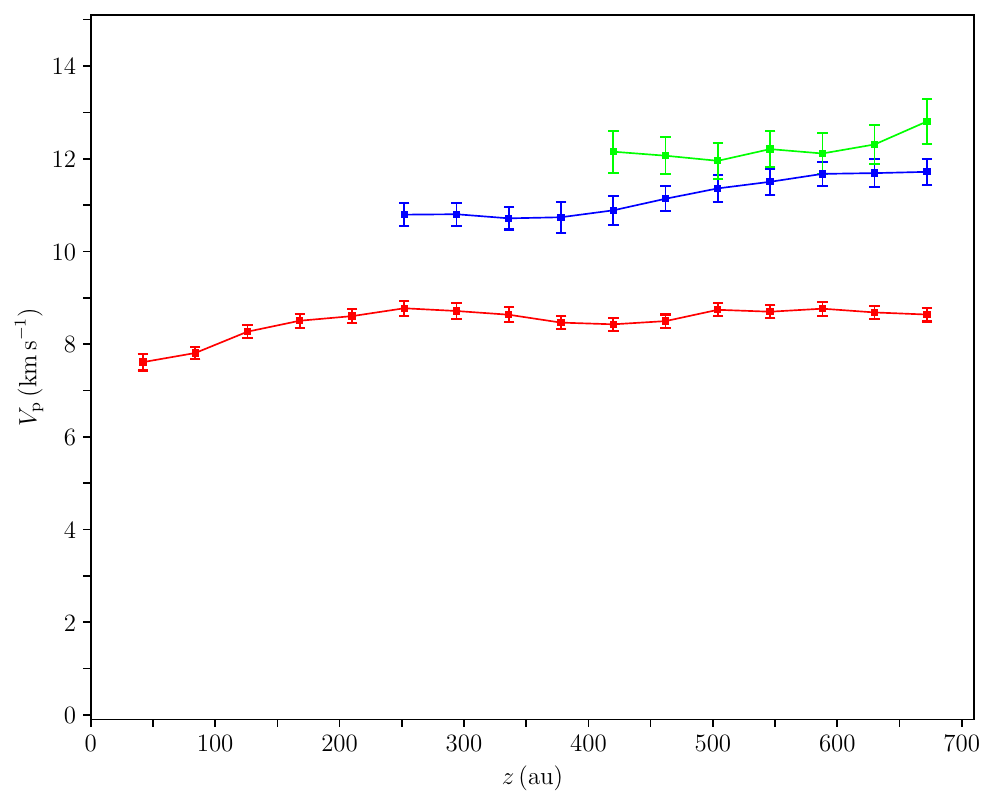}
\caption{\textit{Left panel:} The outward velocity $V_z$ as a function of the height. \textit{Right panel:} poloidal velocity ($V_p=\sqrt{V_r^2+V_z^2}$) as a function of the height. The error bars are derived from the Gaussian fits.}
\label{fig:velocities}
\end{figure*}

The fitted ellipses of Figure \ref{fig:pvmoments} for the shells 1, 2, and 3 are used to determine the kinematic properties shown in Figure \ref{fig:parameters}. The top left panel presents the cylindrical radius as a function of the height for the three shells. These radii follow the general relation of $z=a R^{-\beta/2}$, where the values of $a$ and $\beta$ for the best fit are $a=1.29,\,0.85$, and 0.70, and $\beta=-1.43,\,-1.32$, and -0.92, for the shells 1, 2, and 3, respectively. The top right panel of Figure \ref{fig:parameters} plots the expansion velocity on the line of sight ($V_r\sin i$) as a function of the height. While the shells 1 and 2 reach a constant value $\sim$ 6 $\kms$, the shell 3 presents constant velocity $\sim$4 $\kms$. The rotation velocity on the line of sight ($V_\phi\sin i$) is shown in the bottom left panel. For all shells, the rotation velocity decreases with the height. This behavior is observed in several sources, e.g., Orion Source I (\citealt{Hirota2017}; \citealt{JALV2020}), HH 212 (\citealt{Lee2018}), NGC 1333 IRAS 4C (\citealt{Zhang2018}), and CB 26 (\citealt{JALV2023}; \citealt{Launhardt2023}). It can be observed that the shell 1 has the lowest rotation velocity, while the shell 2 and 3 have the highest rotation velocity, this behavior could be explained if the shells are associated with magnetocentrifugal disk winds, under this assumption, the shell 3 is launching from the innermost region of the accretion disk and the shell 1 is launched from the most extended region.
Finally, the bottom right panel of Figure \ref{fig:parameters} shows the specific angular momentum $j_\mathrm{obs}=R\times V_\phi\sin i$. The specific angular momentum has the same behavior as the rotation velocity, which decreases with height. The shells 2 and 3 seem to have more angular momentum than the shell 1, however, if we consider the error bars in our estimations, we can conclude that the three shells at higher distances from the accretion disk ($z>450$ au), statistically, have the same angular momentum. 
The error bars of these properties are derived through error propagation of the statistical uncertainties extracted in the ellipse fitting plugged into the equations \ref{eq:xoffset}-\ref{eq:rcyl}.

The best fit of the general relation $z=a R^{-\beta/2}$ and the radius of the three shells is compared with measurements of the walls of the molecular outflow in one of the central channels, $9.2\,\kms$, which is shown in the left panel of Figure \ref{fig:channels}. The outflow cavity is observed with more detail in the central channels (close to $V_\mathrm{lsr}=6.9\pm\,0.1\,\kms$) shown in Figures \ref{fig:bluechannels} and \ref{fig:redchannels} in Appendix \ref{sec:apendix_channels}. The radii of the shell 1 (red lines), shell 2 (blue lines), and shell 3 (green lines) trace the outflow cavity until a height of $z<5^{\prime\prime}$, while the shell 3 trace the outflow up to $z\sim20^{\prime\prime}$. The fact that only the shell 3 follows the outflow structure until very high heights could be because the other shells (1 and 2) are older than the shell 3 and the larger expansion and cooling of shells 1 and 2 make them produce fainter CO emission, this effect is observed in perpendicular position-velocity diagrams presented in Figure \ref{fig:zhigh} in Appendix \ref{sec:apendix_channels}.
The right panel of Figure \ref{fig:channels} presents a zoom-in of the inner region of the outflow up to $z\lesssim5^{\prime\prime}$, where our analysis was made.

For the heights considered in our analysis, the variation of the cylindrical radius with the height can be approximated by a cone of semi-opening angle $\tan\theta=R/z$. We fitted these angles and obtain $46.0^\circ\pm0.1^\circ$, $30.0^\circ\pm0.3^\circ$, and $18.8^\circ\pm0.2^\circ$ for the shells 1, 2, and 3, respectively. If the inclination of the cone axis with respect to the line of sight is the same as the inclination of the jet axis, the ratio of the projected velocity components is $\frac{V_r\sin i}{V_\mathrm{cent}}=\tan\theta\times\tan i$. Under this assumption and for $V_{r}\sin i$ and $V_{cent}$ values measurements at each height mentioned above, we estimated the average inclination angle of the jet axis is $87.7\pm 0.3^\circ$.

The velocity $V_z$, by convention, is positive for outward-directed velocity component along the $z$--axis. Once the inclination angle is estimated for the three different shells, we obtain $V_z$ employing eq. (\ref{eq:vz}). The values of this velocity are shown in the left panel of Figure \ref{fig:velocities}. These velocities tend to a constant value for shell 1, while showing a slight increase with height for shells 2 and 3. The highest $V_z$ velocity is presented by shell 3. We also define a poloidal velocity, $V_p=\sqrt{V_r^2+V_z^2}$. As can be seen in the right panel of Figure \ref{fig:velocities}, its dependence with height similar to that of $V_z$.

\begin{figure}[t!]
\centering
\includegraphics[scale=0.525]{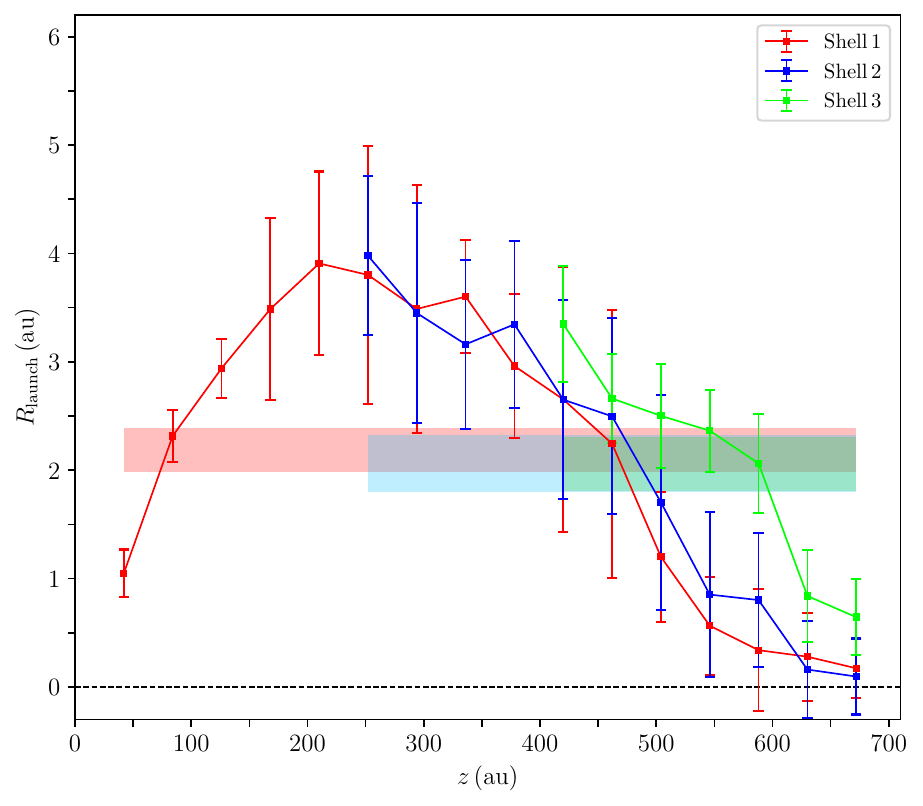}
\caption{The launching radii as a function of the height $z$. These radii are estimated solving the Anderson's relation (\citealt{Anderson2003}). The red, blue, and green rectangles are the mean launching radius for the shell 1, 2, and 3, respectively.
The error bars are derived from the Gaussian fit.}
\label{fig:rlaunching}
\end{figure}

A possibility of the origin of the multiple shell structure is that these shells are launched from different radii from the accretion disk and are driven magnetocentrifugally. Under this scenario, 
we can estimate the launching radius $R_\mathrm{launch}$ of all shells for each height with Anderson's relation \citep{Anderson2003} given by 

\begin{equation}
    \varpi_\infty v_{\phi,\infty}\Omega_0-\frac{3}{2}\left(\mathrm{G}\mstar\right)^{2/3}\Omega_0^{2/3}-\frac{v_{\mathrm{p},\infty}^2+v_{\phi,\infty}^2}{2}\approx0,
    \label{eq:andersonrelation}
\end{equation}
where $\varpi_\infty$ is the distance between the jet axis and the cavity of the molecular outflow, which in our case is determined by the cylindrical radius for each height $z$. The velocities $v_{\phi,\infty}$ and $v_{\mathrm{p},\infty}$ denote the toroidal and the poloidal velocities observed at cylindrical radius, $V_\phi\sin i$ and $V_p$ for this object. The variable G and $\mstar$ are the gravitational constant and the mass of the central protostar (0.45$\pm$0.14 $\msun$). Finally, the variable $\Omega_0$ is the angular speed $\Omega_0=(\mathrm{G}\mstar/\varpi_0^3)^{1/2}$ at launching radius $\varpi_0$. The launching radii for the three different shells as a function of the height are shown in Figure \ref{fig:rlaunching}. 
The derived values of the launching radius are consistent with an outflow origin in the range $0.01<R_\mathrm{launch}<4\,\au$ approximately, this range is consistent with the reported previously by \citet{Louvet2018}. Our estimates for shell 1 present a peculiar behavior, close to the disk ($z\lesssim 200$ au) the launching radius increases with the height, while for the larger distances ($z\gtrsim 200$ au), this radius decreases with the height. Shells 2 and 3 have the same behavior for large distances to the accretion disk, decreasing with height, this behavior could be because the rotation velocity and the specific angular momentum decrease with height as shown in different sources such as Orion Source I (\citealt{Hirota2017}), NGC 1333 IRAS 4C (\citealt{Zhang2018}), HH 212 (\citealt{Lee2018}), and CB 26 (\citealt{JALV2023}; \citealt{Launhardt2023}). However, if we take the mean value (the red, blue, and green rectangles of Figure \ref{fig:rlaunching}), we found that the derived launching region is systematically the same for the three shells.
Nevertheless, since the launching radii of the three shells are $0.01\sim4$ au and they are spatially unresolved, we can summarize that the launching radii of the three shells can be expressed as $2\pm2\,\au$ and we can assume that the three shells are launched from the same region.

\subsection{Mass of the outflow}\label{subsec:massoutflow}

Since the $^{13}$CO is undetected below 3$\sigma$ in the outflow (see also \citealt{Louvet2018}), we can assume that the $^{12}$CO emission is optically thin. Hence, following the formalism in \citet{Zapata2014} we derive a lower limit for the mass of the molecular outflow using:

\begin{eqnarray}
    \left[\frac{\mathrm{M}_{\mathrm{H}_2}}{\msun}\right]&=&1.2\times10^{-15}X_{\frac{\mathrm{H}_2}{\mathrm{CO}}}\left[\frac{\Delta \Omega}{\mathrm{arcsec}^2}\right]\left[\frac{D}{\mathrm{pc}}\right]^2 \nonumber \\
    &\times& \left[\frac{\mathrm{exp}\left(\frac{5.53}{T_{ex}}\right)}{1-\mathrm{exp}\left(\frac{-11.06}{T_{ex}}\right)}\right]\left[\frac{\int I_\nu dv}{\mathrm{Jy}\,\kms}\right],
    \label{eq:moutflow}
\end{eqnarray}
where, we take $X_{\frac{\mathrm{H}_2}{\mathrm{CO}}}=10^{-4}$ for the fractional $^{12}$CO abundance with respect to H$_2$, $\Delta \Omega$ is the solid angle of the source (138 arcsec$^2$), $D$ is the distance (141$\pm$7 pc), $I_\nu$ is the intensity of the emission in jansky, $dv$ is the velocity range in $\kms$, and $T_{ex}$ is the excitation temperature.
 For a excitation temperature of $T_{ex}=30$ K, and under assumption that the emission is governed by a single excitation temperature (\citealt{Louvet2018}) the mass of the outflow is $M_\mathrm{outflow}=1.83\pm0.19\times 10^{-4}$ $\msun$. This value is one order of magnitude bigger than reported by \citet{Louvet2018} and \citet{Pety2006}. 
 The main difference between our mass estimation and their reported mass is the sensitivity. 
 \citet{Louvet2018} consider the emission of the source that exceeds 5$\sigma$ with $\sigma=3.6$ K $\kms$ and an angular size of $\sim7.2$ arcsec$^2$, while in this work, we measured the emission that exceeds 3$\sigma$ with $\sigma=1$ K $\kms$ and an angular size of $\sim138$ arcsec$^2$.

\section{Discussion} \label{sec:discussion}

In this section, we present the different explanations for the multiple shell structure in the molecular outflow of HH 30, and we discuss the magnetocentrifugal process as a possible origin of the molecular outflow. Other possible scenarios such as the photoevaporated disk winds and the entrained material were addressed in the previous work by \citet{Louvet2018}.

\subsection{Origin of the multiple shell structure}\label{subsec:shellstructure}

\begin{figure}[t]
\centering
\includegraphics[scale=0.5]{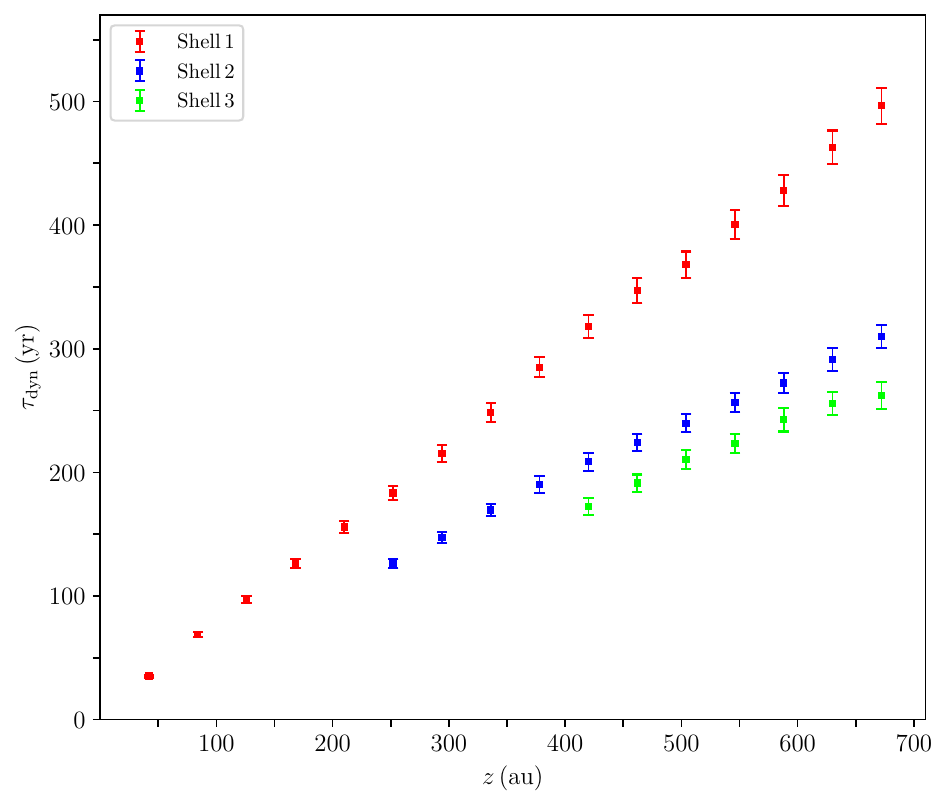}
\caption{Dynamical time of the different shells of the molecular outflow HH 30 as a function of the height $z$. The error bars are derived from the Gaussian fit.}
\label{fig:dynamical}
\end{figure}

The emission of the $^{12}$CO shows that the molecular outflow associated with HH 30 presents the internal structure of the multiple shells as shown in Figure \ref{fig:pvmoments} and Figure 8 of \citet{Louvet2018}. A similar internal structure of the molecular outflow has been reported in several sources such as HH 46/47 (\citealt{Zhang2019}), DO Tauri (\citealt{FernandezLopez2020}), and DG Tau B (\citealt{deValon2022}), where the authors interpreted that the presence of the multiple shells in a molecular outflow is associated with episodic ejections of the material by a wide-angle wind from the accretion disk. To support that the three shells found in the molecular outflow of HH 30 are associated with episodic ejections, we estimated their dynamical time as a function of the height, $\tau_\mathrm{dyn}=z/V_z$. These results are presented in Figure \ref{fig:dynamical}. The shells reach the maximum height ($z_\mathrm{max}=4.8^{\prime \prime}\approx672\,\au$) at $\sim497\pm 15$ yr, $\sim310\pm9$ yr, and $\sim262\pm11$ yr (shell 1, 2, and 3, respectively). Figure \ref{fig:dynamical} shows that the difference in the dynamical ages between the first ejection (shell 1) and the second ejection (shell 2) tends to a constant value of $\sim187\pm17$ yr, the difference of the dynamical age between the second ejection and the third ejection (shell 3) has the same behavior with a value of $\sim 48\pm14$ yr.

We must consider that the estimated dynamical ages of these shells may not be the real age of these components since our estimation does not consider slow-down effects due to the interaction with the surrounding material. Therefore, the dynamical ages of these shells are upper limits of their true age. Under this assumption, the intervals between the different episodic ejections are upper limits too. While these values are much higher than the kinematic ages of the knots $A_{1,2,3}$, B$_{1,2,3}$, and $C$ of the HH 30 jet, they could be associated with the knots $E_{1,2,3\mathrm{b},4}$ located at a height $z>35^{\prime\prime}$ with kinematic ages between $240.8\pm1.7$ yr--$413.7\pm4.9$ yr (\citealt{Estalella2012}). In this case, the ages of our shells 2 and 3 could be consistent with episodic outbursts in the collimated jet, however, the age of shell 1 $\sim$ 500 yr is older than the E knots, this may be because the age of shell 1 is overestimated or the shell 1 could be associated with another episodic outburst has not been observed. This suggest that the episodicity seen in the jet and the outflow may be originated from the same outburst event.

\begin{figure}[t!]
\centering
\includegraphics[scale=0.5]{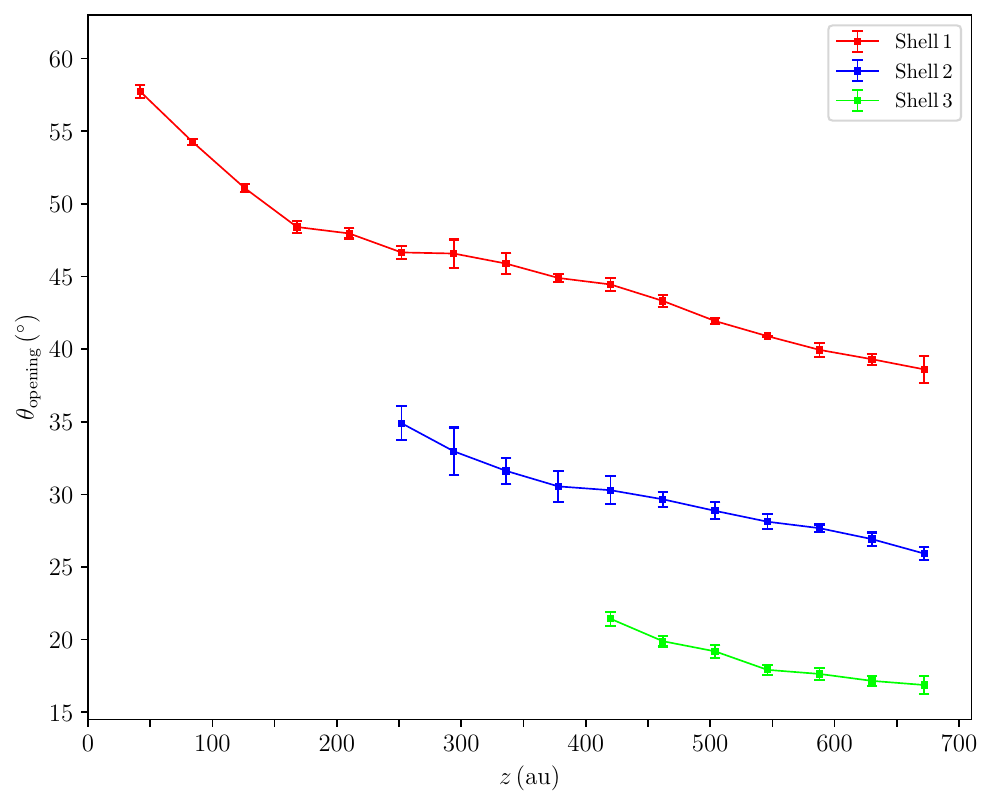}
\caption{Opening angle of the different shells as a function of the height $z$.
The error bars are derived from the Gaussian fit.}
\label{fig:thetaopen}
\end{figure}

The opening angle of the molecular outflow can be an indicator of the evolution of these sources, this is, increases with the source's age. Several studies (e.g., \citealt{Arce2006}; \citealt{Seale2008}; \citealt{Velusamy2014}; \citealt{Hsieh2017}) have shown that this angle widens with time close to the source for different sources. To confirm this assertion, we estimate the opening angle as

\begin{equation}
    \theta_\mathrm{opening}=\tan^{-1}\left(\frac{R-R_\mathrm{launch}}{z}\right)\approx \tan^{-1}\left(\frac{R}{z}\right).
    \label{eq:openingangle}
\end{equation}
The previous approximation takes into account that $R\gg R_\mathrm{launch}$. The difference of this angle with the semi-opening angle mentioned in Section \ref{subsec:kinematicoutflow}, is that this angle are measurement for all heights, while the semi-opening angle mentioned above, is a fit under assumption that the material is following a cone structure. For the three shells, this angle has a maximum value close to the mid-plane at height of $z=0.3^{\prime \prime}$, $\sim57.7^\circ\pm0.5^\circ$ for shell 1, $\sim 34.9^\circ\pm1.2^\circ$ for shell 2, and $\sim21.4^\circ\pm0.5^\circ$ for shell 3 as shown in Figure \ref{fig:thetaopen}. In general, the opening angle will be increasing with time, therefore, the fact that the shell 1 presents the largest opening angle is an indicator that this shell is the oldest, and the shell 3, with the smallest opening angle, is the youngest. These values and this behavior are consistent with produced shells by episodic ejections.

\citet{Shang2023a} through numerical simulations of the x-wind outflows present an alternative explanation for the observed shape of the $^{12}$CO emission of the position-velocity diagrams presented in Figure \ref{fig:pvmoments}. Their model considers that the molecular outflow is the result of the interaction between a wide-angle toroidally magnetized wind with magnetized isothermal toroids that represent molecular cloud cores before the onset of dynamical collapse. Under this assumption, shear, Kelvin--Helmholtz instabilities, and pseudo pulses effects\footnote{The pseudo pulses effects are perturbations in density, poloidal velocity, pressure, and magnetic field strengths produced by oscillations in magnetic forces (\citealt{Shang2020}).} could be responsible for the nested shells observed in the position-velocity diagrams of the molecular outflow of HH 30 for heights $z\leq1.5^{\prime\prime}$. However, provided that the shells are well defined at heights $z>1.5^{\prime \prime}$, the idea that the multiple shell structure is associated with episodic ejections is strengthened. We may assume that the non-detection of the multiple shell structure at heights close to the disk mid-plane ($z\leq1.5^{\prime \prime}$) could be explained by two different ways: first, the observations do not have enough angular resolution to distinguish the multiple shell structure; second, if we assume that the shell 1 is the result of the interaction between the disk wind with its parent cloud, basically as a rotating cloud in gravitational collapse (\citealt{Ulrich1976}), it will tend to stagnate at some point close to the disk mid-plane, because the Ulrich-like envelopes have an infinite density barrier. This barrier would slow-down the shells, allowing for younger shells to catch older ones.

\begin{figure}[t]
\centering
\includegraphics[scale=0.19]{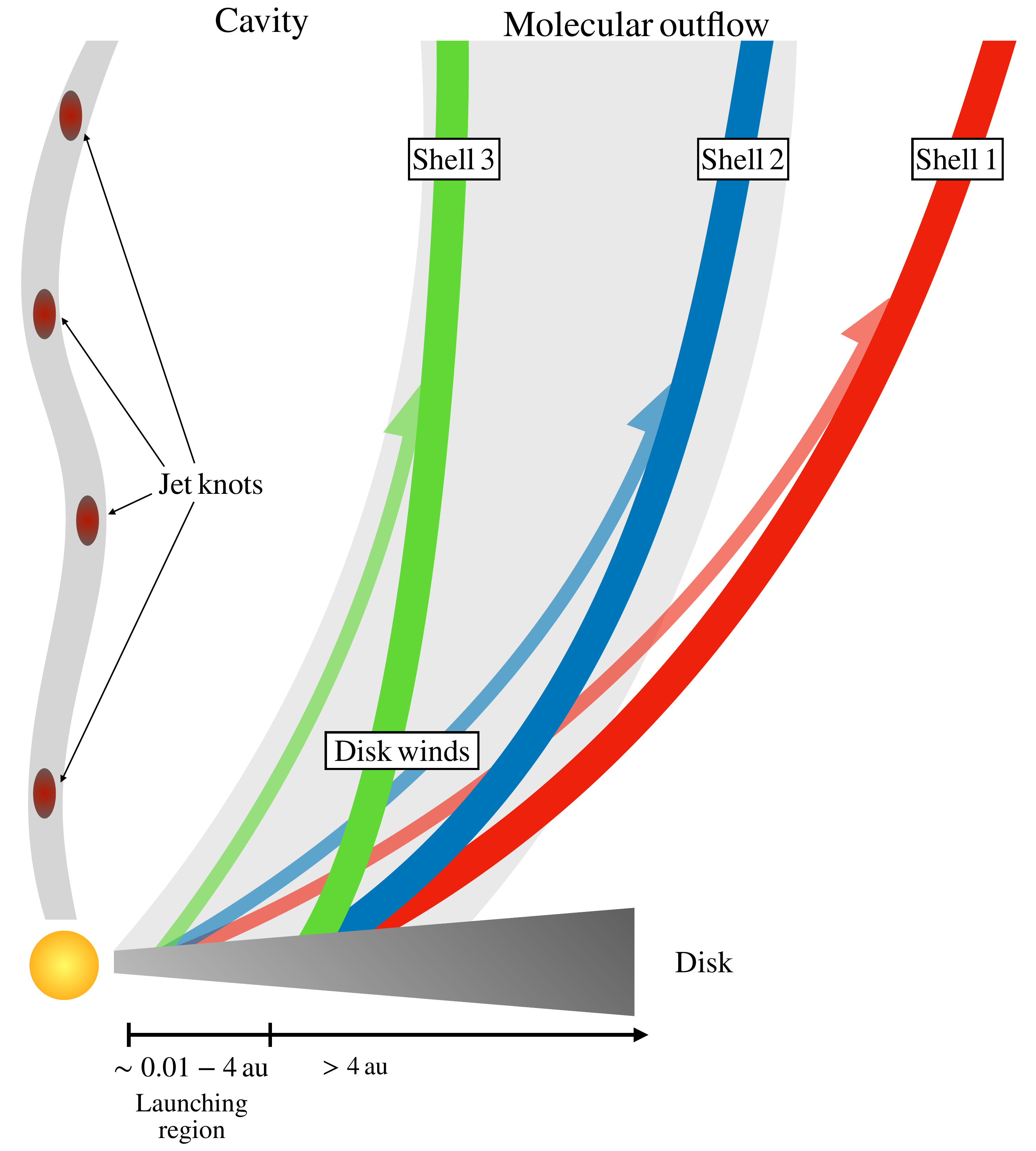}
\caption{Schematic scenario of the different components of source HH 30 under assumption that the outflow is driven by disk winds.}
\label{fig:cartoon}
\end{figure}

\subsection{Origin of the outflow}\label{subsec:originoutflow}

Figures \ref{fig:knots_moments} and \ref{fig:mom0_comp_LHVC} show the structure of the molecular outflow associated with HH 30, in both Figures the southern part of the outflow has not been detected, this may be because the source is located immediately southern boundary of the parental core (e.g., \citealt{Stanke2022}), or this monopolar shape could be consequence of possible deflection effects (e.g., \citealt{FernandezLopez2013}). The origin of this asymmetry is discussed in detail in \citet{Louvet2018}.

We estimated the launching radii of the three shells in section \ref{subsec:kinematicoutflow} through Anderson's relation and we obtain from Figure \ref{fig:rlaunching} that these radii are in a range between 0.01--5 au.
These values are consistent with the expectations for magnetocentrifugal winds of Class II sources (\citealt{Pascucci2022}). In particular, \citet{Anderson2003} assume that the disk winds are driven by magnetic forces with a large-scale poloidal magnetic field anchored in the disk. They also consider that these winds are dynamically cold (negligible enthalpy), axisymmetric, and are in steady state. Therefore, the rotation is governed by the magnetic forces. Hence, wide-angle winds rotate preserving the sense of the rotation of the disk. In the case that these winds were counter-rotating (dynamically warm winds), the launching point could not be estimated using Anderson's relation (\citealt{Tabone2020}).
The HH 30 outflow preserves the direction of the disk rotating, justifying the usage of Anderson's expression to estimate the footpoint of the different shells.

Figure \ref{fig:rlaunching} shows that the mean launching region of shells 1, 2 and 3 is systematically the same. The particular behavior of shell 1 (increases and decreases with height) could be explained if shell 1 is produced by the interaction between the wide-angle disk wind with the surrounding material (e.g., \citealt{JALV2019}). If this is the case, Anderson's relation might not be the best method to estimate the launching point, because this relation considers that the wind has not interacted with the surrounding environment or with itself.

On the other hand, the drastic decrease of the launching radius with the height presented in all shells could be associated with: 1) The poloidal velocity (Figure \ref{fig:velocities}b) of the three shells tends to be constant, therefore, the launching radius only depends on the specific angular momentum behavior, given that this quantity decreases with the height, the launching radius decreases too, however, this could be inconsistent with that expected for a disk wind. 
Hence, this behavior could be an indicator that the two internal shells could be produced by the interaction of the disk wind with itself, and Anderson's relation, the same as with shell 1, is not the best method to estimate their launching radii. 2) Our three shells could be the result of the multiple ejections at three different times but with different launching points associated with the location of the gaps in the accretion disk (e.g., \citealt{Suriano2017}; \citealt{Suriano2018}; \citealt{Suriano2019}). 3) Our measurements of the outward, expansion, and rotation velocities could be contaminated by the entrained material produced by the knots A$_1$ and A$_2$ of the protostellar jet (see Figure \ref{fig:knots_moments}), therefore, our estimation of the launching radii for the shells around those knots is contaminated by this effect too. With our current resolution we can not resolve the disk, and we can not distinguish which effect dominates the behavior of the launching radii as a function of the height, but Anderson's relation could be a good approximation that the launching region of the three shells could be the same as shown in Figure \ref{fig:cartoon}.


\begin{figure}[t]
\centering
\includegraphics[scale=0.5]{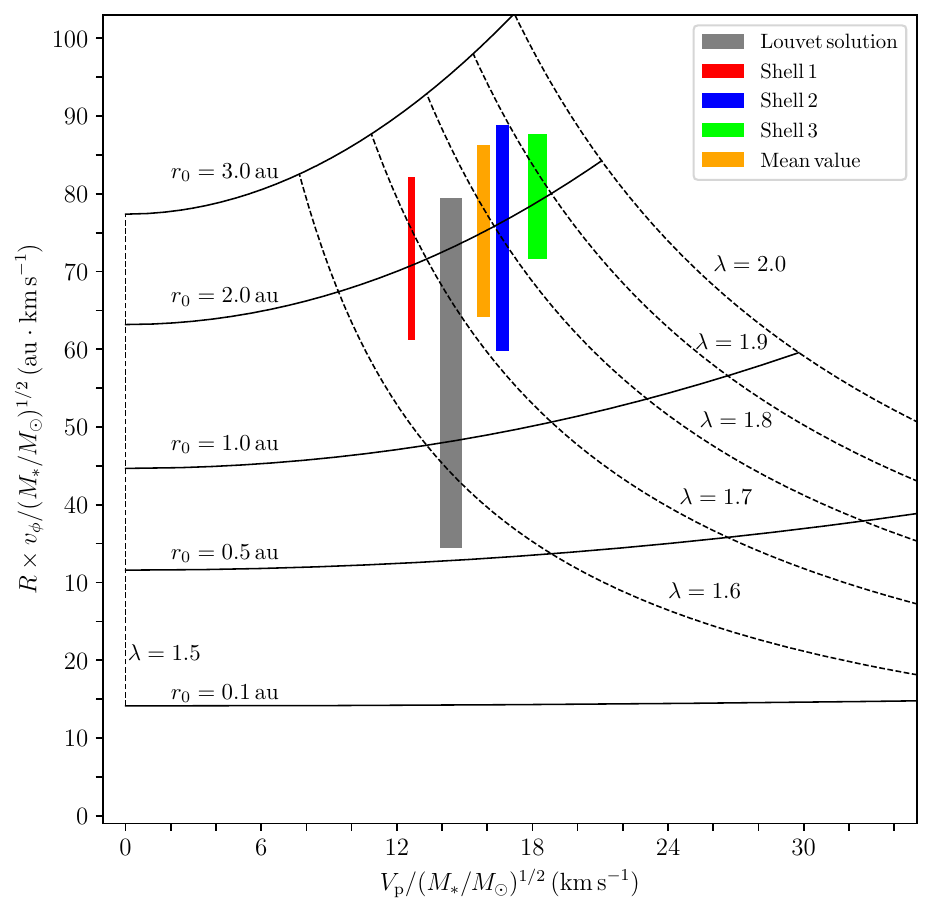}
\caption{Specific angular momentum as a function of the poloidal velocity for steady and axisymmetric MHD disk winds, both normalized to $\sqrt{M_*}$. The black lines represent the expected relation from self-similar cold magneto-centrifugal disk winds with $r_0$ from 0.01 au to 3 au and $\lambda$ from 1.5 to 1.8 (\citealt{Ferreira2006}). The red, blue, and green rectangles show the mean value of the shell 1, shell 2, and shell 3, respectively. While the orange rectangle is the mean value of $\lambda$ of the three shells. The gray rectangle corresponds to the solution for the outflow HH 30 of \citet{Louvet2018}.}
\label{fig:magneticarm}
\end{figure}

The magnetocentrifugal disk winds remove the mass and the angular moment from the accretion disk and exert a torque on the disk surface (e.g., \citealt{Pudritz2007}; \citealt{Alexander2014}; \citealt{Pascucci2022}). An important parameter for describing the magnetic torque is the magnetic lever arm $\lambda=(r_\mathrm{A}/r_0)^2$, where $r_\mathrm{A}$ is the cylindrical radius at the Alfv\'en surface and $r_0$ is the launching point of the streamline that follows the wind. The magnetocentrifugal winds can produce both, very high collimated jets and slow wide-angle disk winds. Their kinematics and morphology depend on the thermal effects on the launching regions. Numerical simulations (e.g., \citealt{Bai2016}) show that for low values of $\lambda$, the magnetocentrifugal winds can extract significant mass and angular momentum from the disk.

As mentioned above, we assume that the three shells of the HH 30 are driven by disk winds associated with three different episodic ejections. Under this assumption, the asymptotic values of the poloidal velocity $V_\mathrm{p}$ and the specific angular momentum for each streamline are (\citealt{Blandford1982}):
\begin{equation}
    V_p=\sqrt{2\lambda-3}\sqrt{G\mstar/r_0},
\end{equation}
\begin{equation}
    R\times v_\phi=\lambda \sqrt{G \mstar r_0}.
\end{equation}
Figure \ref{fig:magneticarm} shows the relation between the specific angular momentum $j=R\times v_\phi$ and the poloidal velocity $V_p$ for the various solutions of the launching point $r_0$ and magnetic lever arm parameter $\lambda$. The mean poloidal velocities and the mean specific angular momentum from the three shells follow the line of $r_0=2\,\au$ and $\lambda\sim1.6-1.9$. For the three shells, the launching radii $r_0$ are consistent with our estimates through Anderson's relation $R_\mathrm{launch}$ and our assumption that the three shells are launched from the same region is confirmed. Also, the low derived limit on $\lambda\sim 1.6-1.9$ is consistent with a solution for warm magnetohydrodynamics disk-wind models (\citealt{Casse2000}) or cold magnetohydrodynamics disk wind models from weakly magnetized disks (\citealt{Jacquemin-Ide2019}). The gray rectangle of Figure \ref{fig:magneticarm} represents the solution derived by \citet{Louvet2018} for HH 30.

In Section \ref{subsec:massoutflow}, we estimated a mass of $1.83 \pm 0.19 \times 10^{-4}\,\msun$ for the HH 30 outflow. For simplicity, we assume that the outward $V_z=12.2\pm0.5\,\kms$, poloidal $V_p=12.8\pm0.5\,\kms$, and rotation $v_\phi=0.13\pm0.04\,\kms$ velocities correspond to the values of the shell 3 at height of $z=4.8^{\prime\prime}\approx 672\,\au$, under assumption that these velocities tend to be a constant at large heights. The size of the molecular outflow is 19$^{\prime\prime}$ with a cylindrical radius of $R=394\pm 20$ au (estimated with the general relation $z=aR^{-\beta/2}$ showed in Section \ref{subsec:kinematicoutflow}). We obtain a dynamical time of $\tau_\mathrm{dyn}=z/V_z=1.04\pm 0.07\times 10^3$ yr, a mass-loss rate of the outflow of $\dot{M}_\mathrm{outflow}=M_\mathrm{outflow}/\tau_\mathrm{dyn}\approx1.76\pm0.21\times10^{-7}\,\msun\,\yr$, a linear momentum rate of $\dot{P}_\mathrm{outflow}=\dot{M}_\mathrm{outflow} V_\mathrm{p}\approx2.25\pm0.29\times10^{-6}\,\msun\,\yr\,\kms$, and a angular momentum rate of $\dot{L}_\mathrm{outflow}=\dot{M}_\mathrm{outflow} R v_\phi=9\pm2.9\times10^{-6}\,\msun\,\mathrm{yr}^{-1}\,\au\,\kms$.

The mass-loss rate of the HH 30 wind is $\dot{M}_w\simeq9\times10^{-8}\,\msun\,\yr$ (\citealt{Louvet2018}), is smaller than the mass-loss rate of the outflow by a factor $\dot{M}_\mathrm{outflow}/\dot{M}_w\sim1.95\pm0.24$. 

If we assume that the outflow is a disk wind $\dot{M}_\mathrm{outflow}=\dot{M}_w\sim f\dot{M}_\mathrm{d,a}$, we can estimate the accretion luminosity at the stellar surface as

\begin{eqnarray}
    L_{a}=\eta \frac{G\mstar \dot{M}_\mathrm{d,a}}{\rstar}\equiv\eta\frac{G\mstar \dot{M}_\mathrm{outflow}}{f \rstar},
\end{eqnarray}
where $\rstar$ is the stellar radius and $\eta\sim0.5$. Using $\mstar=0.45\pm0.14\,\msun$ and $\rstar\sim2-3\,\rsun$, the accretion luminosity is $L_a\geq (1/f)(0.41\pm0.14-0.62\pm0.21)\,\lsun$. This value is consistent with the luminosity of the source of $0.2-0.9\,\lsun$ (\citealt{Cotera2001}) by a factor of $f\sim$0.6--2.

Under a scenario of all mass and angular momentum being removed from the accretion disk by magnetocentrifugal disk winds, the lever arm relates the mass-loss rate with the disk accretion rate as $\dot{M}_w\sim\dot{M}_\mathrm{acc}/\lambda$ (\citealt{Pelletier1992}). This assumption is consistent with the found $\lambda$ values of the three shells and with the found $f$ value for the accretion luminosity.

In summary, the $\lambda$ values found, the estimated rates of the outflow and the disk wind, and the accretion luminosity argue in favor of the scenario with multiple shells driven by a disk wind.

\section{Conclusions} \label{sec:conclusions}
We present a detailed analysis of ALMA archive observations for the molecular line emission of $^{13}$CO and $^{12}$CO from the accretion disk and the molecular outflow, respectively, associated with the protostellar system HH 30. Our main results are the following:
\begin{itemize}
    \item[-] The emission of the $^{13}$CO traces the accretion disk, that presents Keplerian rotation. We estimate the dynamical mass of the central object of the system (central protostar and disk mass) of $M_\mathrm{dyn}= 0.45\pm0.14 \,\msun$.
    \item[-] We identify the internal cavity in the molecular outflow, where the emission of the $^{12}$CO traces the walls of this cavity. Furthermore, the high-velocity of the gas between the S$_{[\mathrm{II}]}$ knots of the precessing jet could be a probe that the molecular outflow is a combination of the entrained material by the jet and the disk winds launched directly from the accretion disk.
    \item[-] The perpendicular position-velocity diagrams to the jet axis show a structure with multiple internal shells. We detect three different shells associated with the episodic ejections of a wide-angle wind from the accretion disk. The dynamical times of the shells are $\sim 497\pm 15$ yr (shell 1), $\sim 310\pm9$ yr (shell 2), and $\sim 262\pm11$ yr (shell 3). The difference between the first and the second events is $\sim187\pm17$ yr, and $\sim 48\pm14$ yr between the second and the last events.
    \item[-] The kinematics of the different shells show that the three shells are in constant expansion in the radial direction and present signatures of rotation.
    \item[-] Our estimations of the launching radii $2\pm2$ au and the magnetic lever arm $\lambda\sim 1.6-1.9$ of the three shells are consistent with the expected values if the molecular outflow is launched through magnetocentrifugal processes.
    \item[-] The lower limit of the mass of the molecular outflow is $M_\mathrm{outflow}=1.83\pm0.19\times10^{-4}\,\msun$, with a mass--loss rate of $1.76\pm0.21\times10^{-7}\,\msun\,\yr$, linear momentum rate of $2.25\pm0.29\times10{-6}\,\msun\,\yr\,\kms$ and an angular rate of $9\pm2.9\times10{-6}\,\msun\,\yr\,\kms\,\au$. As a result of the comparison of these rates with the mass, linear and angular momentum of the wind, we obtain that these rates are very similar. We also found that the accretion luminosity is consistent with the luminosity of the central source by a factor of $f\sim$0.6--2. 
    \item[-] The dynamical times, the launching radii, and the magnetic lever arm of the three shells, as well as, the mass, the linear and the angular momentum rates of the outflow, are strong evidence that the molecular outflow associated with HH 30 system is originated by episodic ejections of a slow wide-angle disk wind.
\end{itemize}

\begin{acknowledgments}
We thank an anonymous referee for very useful suggestions that improved the presentation of this paper.
J. A. L\'opez-V\'azquez and Chin--Fei Lee acknowledge grants from the National Science and Technology Council of Taiwan (NSTC 110--2112--M--001--021--MY3 and 112--2112--M--001--039--MY3) and the Academia Sinica (Investigator Award AS--IA--108--M01). L. A. Z. acknowledges financial support from CONACyT-280775 and UNAM-PAPIIT IN110618, and IN112323 grants, M\'exico. This paper makes use of the following ALMA data: ADS/JAO.ALMA\#2013.1.01175.S and ADS/JAO.ALMA\#2018.1.01532.S. ALMA is a partnership of ESO (representing its member states), NSF (USA) and NINS (Japan), together with NRC (Canada), MOST and ASIAA (Taiwan), and KASI (Republic of Korea), in cooperation with the Republic of Chile. The Joint ALMA Observatory is operated by ESO, AUI/NRAO and NAOJ.
\end{acknowledgments}


\appendix
\section{Supplementary figures}
\label{sec:apendix_channels}

\begin{figure*}[h]
\centering
\includegraphics[scale=0.36]{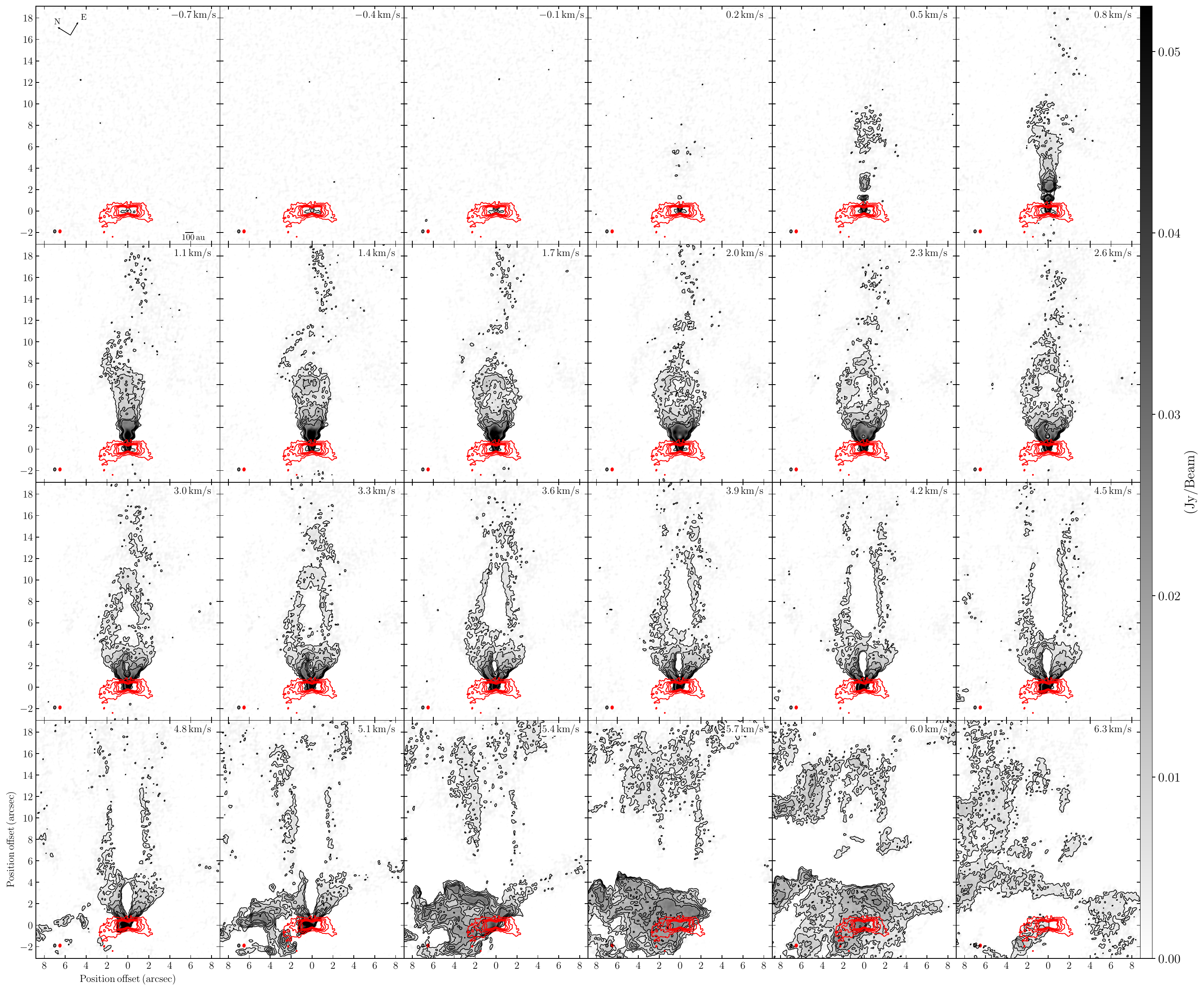}
\caption{Blueshifted channel maps of the $^{12}$CO molecular line emission of the molecular outflow of HH 30. The channel velocity is indicated in the right top. The contours levels start at 3$\sigma$ in steps of 5$\sigma$, 10$\sigma$, 15$\sigma$, and 20$\sigma$, where $\sigma=0.96$ Jy/Beam. The red contours corresponds to the moment zero (integrated intensity) of the HH 30 disk emission of $^{13}$CO molecular line, the contours levels start at 5$\sigma$ in steps of 5$\sigma$, 10$\sigma$, 15$\sigma$, and 20$\sigma$, where $\sigma=1.07\times10^{-3}$ Jy/Beam. The synthesized beams in all panels are shown in the lower left corner.}
\label{fig:bluechannels}
\end{figure*}

\begin{figure*}[t]
\centering
\includegraphics[scale=0.36]{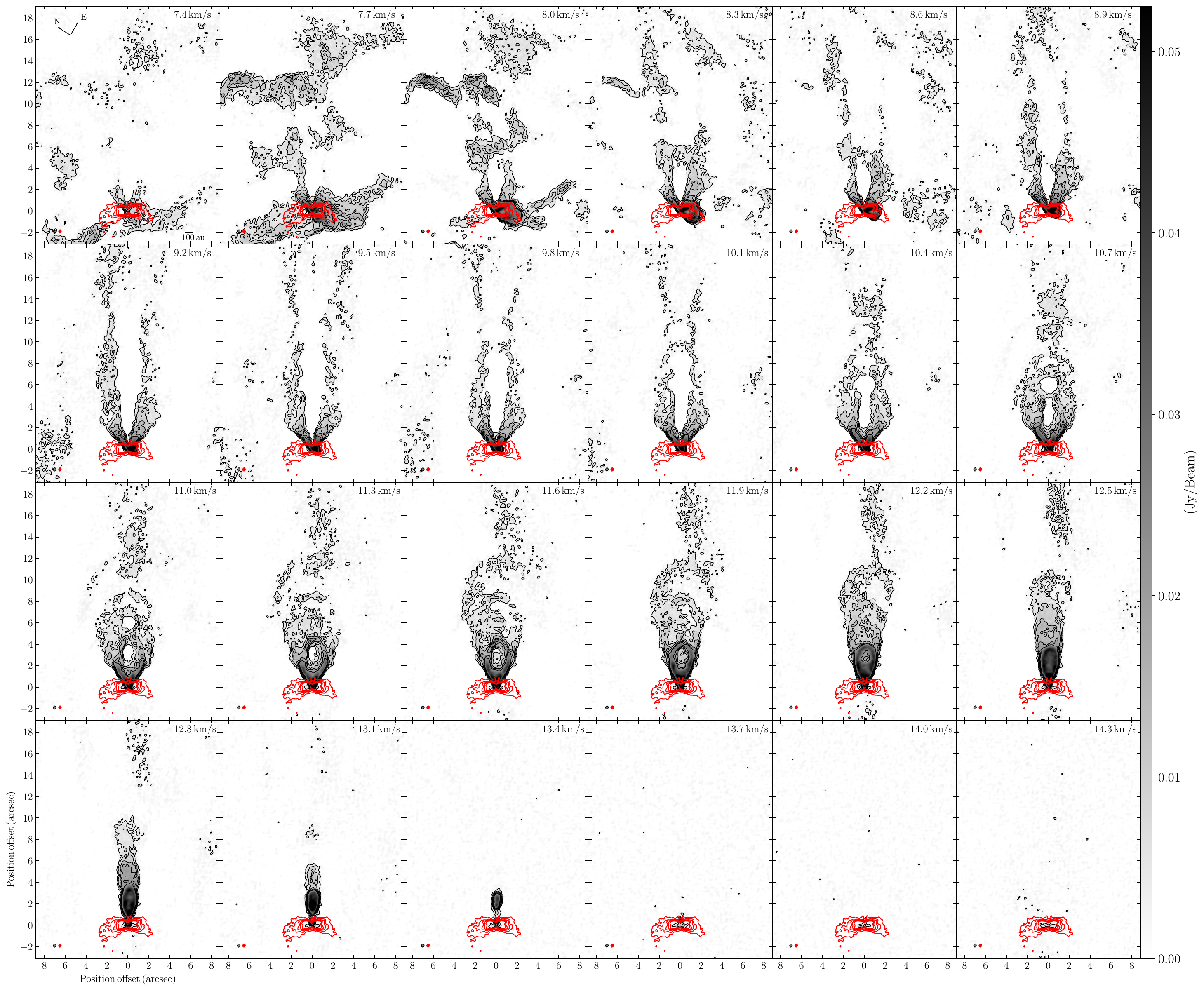}
\caption{Redshifted channel maps of the $^{12}$CO molecular line emission of the molecular outflow of HH 30. The same description as Figure \ref{fig:bluechannels}.}
\label{fig:redchannels}
\end{figure*}

\begin{figure*}[t]
\centering
\includegraphics[scale=0.465]{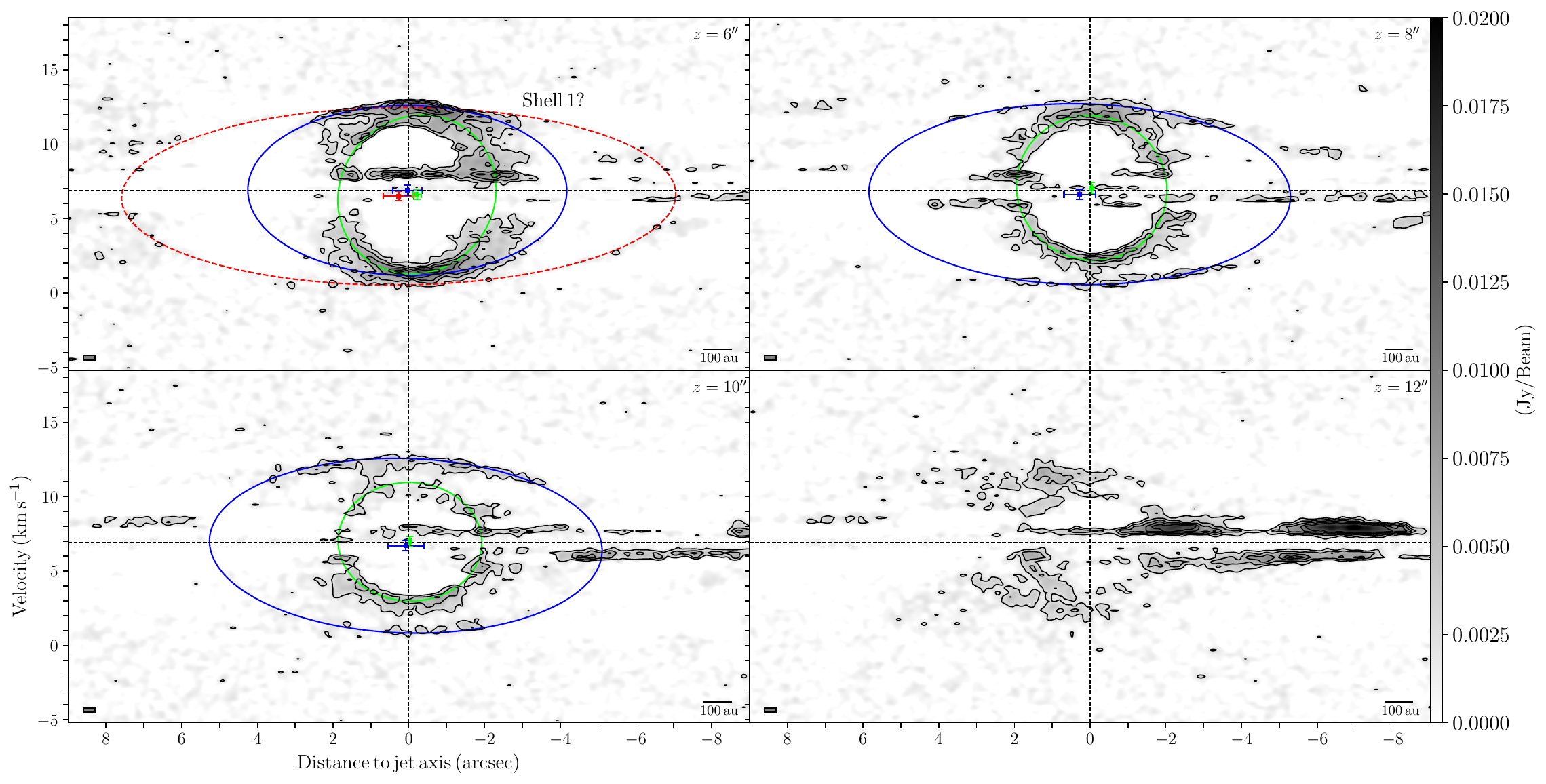}
\caption{Position-velocity diagrams from $^{12}$CO emission perpendicular to jet axis at different heights from $z=6^{\prime\prime}$ (840 au) to $z=12^{\prime\prime}$ (1680 au) with an interval of 2$^{\prime\prime}$ (280 au). The same description as Figure \ref{fig:pvmoments}.}
\label{fig:zhigh}
\end{figure*}



\end{document}